\documentclass[preprint,12pt]{elsarticle}
\usepackage{amssymb}
\usepackage{braket}
\usepackage{epstopdf}
\usepackage{lscape}
\journal{Nuclear Physics A}
\begin{document}
\begin{frontmatter}
\title{Unique first-forbidden $\beta$-decay transitions in odd-odd and even-even heavy nuclei}
\author{Jameel-Un Nabi$^{1}$, Necla \c{C}akmak$^{2}$, Muhammad Majid$^{1}$ and Cevad Selam$^{3}$}
\address{$^{1}$Faculty of Engineering Sciences, GIK Institute of Engineering Sciences and Technology, Topi 23640, Khyber Pakhtunkhwa, Pakistan}
\address{$^{2}$Department of Physics, Karab\"uk University, 78050, Karab\"uk,Turkey}
\address{$^{3}$Department of Physics, Mu\c{s} Alparslan University, 49000, Mu\c{s},Turkey}
\begin{abstract}
The allowed Gamow-Teller (GT) transitions are the most common weak
nuclear processes of spin-isospin $(\sigma\tau)$ type. These
transitions play a key role in numerous processes in the domain of
nuclear physics. Equally important is their contribution in
astrophysics, particularly in nuclear synthesis and
supernova-explosions. In situations where allowed GT transitions are
not favored, first-forbidden transitions become significant,
specifically in medium heavy and heavy nuclei. For neutron-rich
nuclei, first-forbidden transitions are favored mainly due to the
phase-space amplification for these transitions. In this work we
calculate the allowed GT as well as unique first-forbidden (U1F)
$|\Delta$J$|$ = 2 transitions strength in odd-odd and even-even
nuclei in mass range $70\leq A \leq214$. Two different pn-QRPA
models were used with a schematic separable interaction to calculate
GT and U1F transitions. The inclusion of U1F strength improved the
overall comparison of calculated terrestrial $\beta$-decay
half-lives in both models. The \textit{ft} values and reduced
transition probabilities for the $2^-\longleftrightarrow 0^+$
transitions were also calculated. We compared our calculations with
the previously reported correlated RPA calculation and experimental
results. Our calculations are in better agreement with measured
data. For stellar applications we further calculated the allowed GT
and U1F weak rates. These include $\beta^{\pm}$-decay rates and
electron/positron capture rates of heavy nuclei in stellar matter.
Our study shows that positron and electron capture rates command the
total weak rates of these heavy nuclei at high stellar temperatures.

\end{abstract}
\end{frontmatter}
 \section{Introduction}
 The connection between the generation of energy in stars and weak
interaction leads to large-scale stellar events. Supernova
explosions and related physics are one of the most studied phenomena
to be known in astrophysics. Supernovae are intimately connected
with the nucleosyntheis problem (e.g. \cite{Cow04}. The weak decay
processes are the crucial constituents in all major astrophysical
events. The core of the massive star collapses due to  weak
interaction reactions activating a supernova explosion. Another key
phenomenon where weak interactions play a significant part includes
neutronisation of stellar core via capturing of electrons by nuclei
and by free protons. This process effects the creation of heavier
elements beyond  iron through $r$-process during late phases of
evolution of massive stars. The weak rates determine the mass of the
core and provide a fair estimate of the fate and strength of the
shock wave produced by the supernova outburst \cite{Bor06,Nabi10}.
The $\beta$-decay properties of neutron rich nuclei are essential in
order to understand the $r$-process. Though in astrophysical context
the site of the $r$-process is not known with certainty, it is
normally accepted that it takes place in an explosive environment
possessing extremely high neutron densities ($\ge 10^{20}$
cm$^{-3}$) and high temperatures (T $\ge 10^{9}$ K). Under these
conditions, neutrons are captured more rapidly than the competing
$\beta$-decays and in the nuclear chart the $r$-process path goes
through the neutron-rich domain with comparatively small (and having
round-about constant) neutron separation energies  of $\lesssim $ 3
MeV \cite{Zhi13}.

The theoretical explanation of weak processes, especially the double
$\beta$-decay process, is an open question for the nuclear-structure
theories and it has significance to explore physics beyond the
standard model \cite{Civ96}. The compiled theoretical and
experimental results can be seen in \cite{Tre95}. The analysis of
some of the recent developments, both in experiments and theory were
discussed in \cite{Moe94}. The estimation of $\beta$-decay
half-lives, in agreement with the experimental values, is one of the
challenging difficulties for nuclear theorists. The $\beta$-decay
rates and half-lives of nuclei are determined widely by using
various nuclear models. Takahashi \textit{et al.} \cite{Tak69}
calculated these rates using the  gross theory which is statistical
in nature. In this model the shell structure of nucleons is not
entirely accounted for and the theory takes into account the average
values of $\beta$-strength functions. Very soon it was realized to
use a microscopic model for a more reliable calculation of
$\beta$-decay half-lives. In studies of nuclear $ \beta$-decay
properties the proton-neutron quasi particle random phase
approximation (pn-QRPA) theory has been widely used in literature.
In pn-QRPA a quasiparticle basis via pairing interaction is
constructed first, and then the equation of RPA having schematic
Gamow-Teller (GT) residual interaction is solved. Sorensen and
Halbleib \cite{Hal67} developed this model by simplifying the usual
RPA to calculate the relevant transitions. The pn-QRPA calculations
were then extended to deformed nuclei by many authors
\cite{Ran73,Gab73,Gab76,Iva77,Mut89,Mol90}. Microscopic calculations
of allowed weak rates, from atomic number 6 to 114, were first
performed by \cite{Kla84}. Later the pn-QRPA model was used to
calculate $ \beta$-decay properties of nuclei far from line of
stability both in the $\beta$-decay \cite{Sta90} and electron
capture direction \cite{Hir93}. These calculations
\cite{Sta90,Hir93} highlighted the strong and reliable predictive
power of the pn-QRPA model specially as one moves far from the line
of stability. The pn-QRPA model was later used for calculation of
unique first-forbidden (U1F) transitions ($|\Delta$J$|$ = 2)  by
\cite {Hom96,Ken05} under terrestrial conditions. This calculation
demonstrated that for near-magic and near-stable nuclei, greater
contribution to the total transition strength came from U1F
transitions (see Fig.~ 9 and Table IX of \cite {Hom96}). It was Nabi
and Klapdor-Kleingrothaus who used the pn-QRPA model, for the first
time, to calculate weak interaction rates in stellar matter
\cite{Nab99a, Nab99, Nab04}. The same model was later modified to
calculate U1F rates in stellar matter by Nabi and Stoica
\cite{Nab14}. The QRPA studies based on the Fayans energy functional
was extended by Borzov for a consistent treatment of allowed and
first-forbidden (FF) contributions to $r$-process half-lives
\cite{Bor06}. In \cite{Bor06} it was shown that the first forbidden
transitions contribute dominantly to the total weak decay half-life,
mostly for the nuclei having closed Z and N shells. More recently
the authors of Ref. \cite{Zhi13} calculated the half-lives for
$r$-process waiting-point nuclei using the large scale shell-model
including FF contributions.

There are several experimental and theoretical results available
about the allowed transitions but literature is rather scarce when
it comes to discussion of FF transitions . Suhonen \cite{Suh93}
studied the electron capture/$\beta^{+}$ transitions for $^{136}La$
($1^{+}$)$\rightarrow$ $^{136}Ba$ ($J^{\pi}$) and  $\beta$-decay
properties for $^{136}Cs$($5^{+}$)$\rightarrow$
$^{136}Ba$($J^{\pi}$) and $^{136}I$ ($2^{-}$) $\rightarrow$
$^{136}Xe$ ($J^\pi$). The $\beta$-decay strength were calculated in
allowed and FF approximations including only the ground state
transitions. The $\beta$-decay properties for odd-odd nuclei to the
excited levels of neighboring even-even nuclei were discussed in
detail within the QRPA theory. A common vacuum was assumed and
harmonic oscillator basis were employed in this calculation
\cite{Suh93}. The effects of spin-isospin dependent interactions on
even-even and odd-odd nuclei and FF beta decay transitions for
$|\Delta$J$|$ = 0, 2 were studied by Civitarese \emph{et al.}
\cite{Civ86} using RPA technique for two different quasiparticle
excitations. Here the authors assumed that the relativistic
$\beta$-moment $M^{\pm}(\rho_{A}, \lambda=0)$ is proportional to the
matrix element of non-relativistic one. For
$0^{+}\longleftrightarrow 0^{-}$ the experimental and calculated
$ft$ values were in better agreement. For U1F transitions the
renormalization effects improved the theoretical values (see Fig.~3
of \cite{Civ86}). The $0^{+}\longleftrightarrow 0^{-}$ FF beta
transitions for some spherical nuclei in the mass region $90\leq A
\leq214$ were later studied by \c{C}akmak \emph{et al.} \cite{Nec10}
using the pn-QRPA method. In this work the relativistic
$\beta$-moment matrix elements were calculated directly and without
any assumption. The results obtained by \cite{Nec10} were in better
agreement with the experimental data and previous calculations.
Using the shell-model, the $\beta$-decay half-lives of N = 126
isotones have been calculated by Suzuki \emph{et al.} \cite{Suz12},
considering both the GT and FF transitions. It was found that the FF
transitions reduce the half-lives, by nearly twice to several times,
from those by the GT contributions only. Recently the $\beta$-decay
half-lives and $\beta$-delayed neutron emission probabilities
(P$_{n}$) for 5409 nuclei have been studied by \cite{Mar16} by using
the pn-RQRPA model. It was concluded that FF transitions contribute
significantly to the total decay rate and must be taken into account
regularly in modern evaluations of half-lives.

In this paper we calculate U1F $\beta$-decay transitions for both
spherical and deformed odd-odd and even-even nuclei using two
different pn-QRPA models, having separable residual GT interactions.
Earlier rank 0 ($0^{+}\longleftrightarrow 0^{-}$) FF transitions for
these heavy nuclei were studied by \c{C}akmak \emph{et al.}
\cite{Nec10} using the pn-QRPA(WS) method. We are currently working
on the calculation of rank 0 FF transitions using the pn-QRPA(N)
model and hope to report this in future. Accordingly we restrict
ourselves to calculations of U1F $\beta$-decay transitions in this
paper. For the first pn-QRPA(WS) calculation, we employ the
Woods-Saxon potential with Chepurnov parametrization \cite{Sol76}.
Only the $ph$ term of $\beta$- decay effective interaction was
considered for calculation of U1F transitions. The pairing
correlation constants were taken as $C_{n}=C_{p}= 12/\sqrt{A}$ for
open shell nuclei. The strength parameter of the effective
interaction was taken as $\chi_{U1F}=350A^{-5/3}~MeVfm^{-2}$. The
second pn-QRPA(N) model used a deformed Nilsson potential. Further
GT interaction was explored both in the particle-hole ($ph$) and
particle-particle ($pp$) channels in the pn-QRPA(N) model. Using the
Nilsson basis, the strength parameters of the $ph$ and $pp$ forces
were chosen so as to best reproduce the values of experimental
$\beta$-decay half-lives. In this project the effective ratio of
axial and vector coupling constants (g$_A$/g$_V$)$^{2}$$_{eff}$ was
taken as: (g$_A$/g$_V$)$^{2}$$_{eff}$ =
0.7(g$_A$/g$_V$)$^{2}$$_{bare}$, with (g$_A$/g$_V$)$^{2}$$_{bare}$ =
-1.254. The same quenching factor of 0.7 was also used in
\cite{Civ86}.

The theoretical description of allowed and U1F transitions using the
pn-QRPA(WS) and pn-QRPA(N) models are discussed in next section. The
pn-QRPA(N) model was later used to calculate allowed GT and U1F
rates in stellar matter. We are currently working on calculation of
stellar rates using the pn-QRPA(WS) model and this would be treated
as a future assignment. In Sec.~3 we discuss the results of our
calculations and also compare them with experimental data and other
theoretical models. We finally conclude our findings in Sec.~4.

\section{Theoretical Formalism}
As mentioned earlier, for calculation of allowed GT and U1F
($|\Delta$J$|$ = 2) transitions, two different pn-QRPA models were
considered in this work. The first model i.e. pn-QRPA(WS) considered
only the spherical nuclei, in which the Woods-Saxon potential basis
was used and the U1F transitions were calculated. The second pn-QRPA
model is mentioned as pn-QRPA(N) in this paper. A separable GT force
with $ph$- and $pp$-channels was used in order to reduce the
 eigenvalue equation to a fourth order algebraic equation
(for details see \cite{Mut92}) which can be solved with much ease
and less computation time. In the pn-QRPA(N) model deformation of
nuclei was considered. Allowed GT and U1F transitions, both
terrestrial and stellar, were calculated within the pn-QRPA(N)
theory. In the next sub sections the formalism for pn-QRPA(WS) and
pn-QRPA(N) models is briefly described.

\subsection{Model Hamiltonian in the  pn-QRPA(WS) model}
The Hamiltonian which produces the spin-isospin-dependent
vibrational modes (rank 2) in odd-odd nuclei within pn-QRPA(WS)
model is specified by
\begin{eqnarray}
\hat{H}=\hat{H}_{sqp}+\hat{h}_{ph},
\end{eqnarray}
the single quasi-particle ($sqp$) Hamiltonian of the system is given
as follows
\begin{eqnarray}
\hat{H}_{sqp}=\sum_{j_{\tau}}\varepsilon_{j_{\tau}}\alpha^{+}_{j_{\tau}m_{\tau}}\alpha_{j_{\tau}m_{\tau}}~~~~~~(\tau=p,n),
\end{eqnarray}
where $\varepsilon_{j_{\tau}}$ and $\alpha^{+}_{j_{\tau}m_{\tau}}
(\alpha_{j_{\tau}m_{\tau}})$ represent the nucleons sqp energy and
the quasi-particle creation(annihilation) operators, respectively.
The $\hat{h}_{ph}$ is the spin-isospin effective interaction for U1F
transitions in particle-hole ($ph$) channel and generally given as
\begin{eqnarray}
\hat{h}_{ph}=\frac{2\chi_{2}}{g_{A}}\sum_{j_{p}j_{n}j_{p'}j_{n'}\mu}\{b_{j_{p}j_{n}}A^{+}_{j_{p}j_{n}}(\lambda\mu)+(-1)^{\lambda-\mu}\bar{b}_{j_{p}j_{n}}A_{j_{p}j_{n}}(\lambda-\mu)\}\times\nonumber
\end{eqnarray}
\begin{eqnarray}
\{b_{j_{p'}j_{n'}}A_{j_{p'}j_{n'}}(\lambda\mu)+(-1)^{\lambda-\mu}\bar{b}_{j_{p'}j_{n'}}A^{+}_{j_{p'}j_{n'}}(\lambda-\mu)\},\nonumber
\end{eqnarray}
where $\chi_{2}$ is the ph effective interaction constant.
$A^{+}_{j_{p}j_{n}}(\lambda\mu)$ and $A_{j_{p}j_{n}}(\lambda\mu)$ are the quasi-boson creation and annihilation operators and given by
\begin{eqnarray}
A^{+}_{j_{p}j_{n}}(\lambda\mu)=\sqrt{\frac{2\lambda+1}{2j_{p}+1}}\sum_{m_{p}m_{n}}(-1)^{j_{n}-m_{n}}\langle j_{n}m_{n}\lambda\mu|j_{p}m_{p}\rangle\alpha^{+}_{j_{p}m_{p}}\alpha^{+}_{j_{n}-m_{n}},\nonumber
\end{eqnarray}
\begin{eqnarray}
\{A^{+}_{j_{p}j_{n}}(\lambda\mu)\}^{\dag}=A_{j_{p}j_{n}}(\lambda\mu).\nonumber
\end{eqnarray}
The $b_{j_{p}j_{n}}$, $\bar{b}_{j_{p}j_{n}}$ are the reduced matrix
elements of the non-relativistic multipole operators and defined by
\begin{eqnarray}
b_{j_{p}j_{n}}=<j_{p}(l_{p}s_{p})\|r_{k}\{{Y_{1}(r_{k})\sigma(k)}\}_{2\mu}\|j_{n}(l_{n}s_{n})>V_{j_{n}}U_{j_{p}},\nonumber
\end{eqnarray}
\begin{eqnarray}
\bar{b}_{j_{p}j_{n}}=<j_{p}(l_{p}s_{p})\|r_{k}\{{Y_{1}(r_{k})\sigma(k)}\}_{2\mu}\|j_{n}(l_{n}s_{n})>U_{j_{n}}V_{j_{p}},\nonumber
\end{eqnarray}
where $U_{j_{p}}(U_{j_{n}})$ and $V_{j_{p}}(V_{j_{n}})$ are the
standard BCS occupation amplitudes. The Hamiltonian Eq.~(1) can be
linearized in the pn-QRPA(WS) model. Hence charge-exchange $2^{-}$
vibration modes in odd-odd nuclei are considered as phonon
excitations and are defined by
\begin{eqnarray}
|\Psi_{i}\rangle=Q^{+}_{i}(\mu)|0\rangle=\sum_{j_{p}j_{n}\mu}\{\psi^{i}_{j_{p}j_{n}}(\mu)A^{+}_{j_{p}j_{n}}(\lambda\mu)-\varphi^{i}_{j_{p}j_{n}}(\mu)A_{j_{p}j_{n}}(\lambda\mu)\}|0\rangle,\nonumber
\end{eqnarray}
where $Q^{+}_{i}(\mu)$ is the pn-QRPA phonon creation operator,
$|0\rangle$ is the phonon vacuum which corresponds to the ground
state of an even-even nucleus and performs $Q_{i}(\mu)|0\rangle=0$
for all $\emph{i}$. The $\psi^{i}_{j_{p}j_{n}}(\mu)$ and
$\varphi^{i}_{j_{p}j_{n}}(\mu)$ are the forward and backward
quasi-boson amplitudes, respectively. The phonon operators satisfy
the commutation relations
\begin{eqnarray}
\langle0|[Q_{i}(\mu), Q^{+}_{j}(\mu')]|0\rangle=\delta_{ij}\delta_{\mu\mu'}~~~~and~~~~\langle0|[Q_{i}(\mu), Q_{j}(\mu')]|0\rangle=0.\nonumber
\end{eqnarray}
The quasi-boson amplitudes $\psi^{i}_{j_{p}j_{n}}(\mu)$ and
$\varphi^{i}_{j_{p}j_{n}}(\mu)$ satisfy the orthonormalization
condition

\begin{eqnarray}
\sum_{j_{p}j_{n}\mu\mu'}\{\psi^{i}_{j_{p}j_{n}}(\mu)\psi^{i'}_{j_{p}j_{n}}(\mu')-\varphi^{i}_{j_{p}j_{n}}(\mu)\varphi^{i'}_{j_{p}j_{n}}(\mu')\}=\delta_{ii'}\delta_{\mu\mu'}.
\end{eqnarray}
Solving the equation of motion
\begin{eqnarray}
[H,Q^{+}_{i}(\mu)]|0\rangle=\omega_{i}Q^{+}_{i}(\mu)|0\rangle,
\end{eqnarray}
where $\omega_{i}$ is the \emph{i}th $2^{-}$ excitation energy in odd-odd nuclei calculated from the ground state of the parent even-even nucleus. The pn-QRPA(WS) equations take the form
\begin{eqnarray}
\sum_{j_{p}j_{n}j_{p'}j_{n'}\mu}\{\rho_{j_{p}j_{n}j_{p'}j_{n'}}\psi^{i}_{j_{p}j_{n}}(\mu)-\eta_{j_{p}j_{n}j_{p'}j_{n'}}\varphi^{i}_{j_{p}j_{n}}(\mu)\}=\omega_{i}\psi^{i}_{j_{p}j_{n}}(\mu),
\end{eqnarray}
\begin{eqnarray}
\sum_{j_{p}j_{n}j_{p'}j_{n'}\mu}\{\eta_{j_{p}j_{n}j_{p'}j_{n'}}\psi^{i}_{j_{p}j_{n}}(\mu)-\rho_{j_{p}j_{n}j_{p'}j_{n'}}\varphi^{i}_{j_{p}j_{n}}(\mu)\}=\omega_{i}\varphi^{i}_{j_{p}j_{n}}(\mu).
\end{eqnarray}
Here $\rho_{j_{p}j_{n}j_{p'}j_{n'}}$ and
$\eta_{j_{p}j_{n}j_{p'}j_{n'}}$ are the pn-QRPA(WS) matrices
\begin{eqnarray}
\rho_{j_{p}j_{n}j_{p'}j_{n'}}=E_{j_{p}j_{n}}\delta_{j_{n}j_{n'}}\delta_{j_{p}j_{p'}}+2\chi_{2}\{b_{j_{p}j_{n}}b_{j_{p'}j_{n'}}+\bar{b}_{j_{p}j_{n}}\bar{b}_{j_{p'}j_{n'}}\},\nonumber
\end{eqnarray}
\begin{eqnarray}
\eta_{j_{p}j_{n}j_{p'}j_{n'}}=- 2\chi_{2}(-1)^{\lambda-\mu} \{b_{j_{p}j_{n}}\bar{b}_{j_{p'}j_{n'}}+ b_{j_{p'}j_{n'}}\bar{b}_{j_{p}j_{n}}\},\nonumber
\end{eqnarray}
where $E_{j_{p}j_{n}}=\varepsilon_{j_{n}}+\varepsilon_{j_{p}}$ is
the single particle energy. Excitation energies $\omega_{i}$ and
$\psi^{i}_{j_{p}j_{n}}(\mu)$, $\varphi^{i}_{j_{p}j_{n}}(\mu)$
amplitudes are found from Eqs. (3), (5) and (6).

\subsection{$\beta$- decay operator and transition probability in the pn-QRPA(WS) model}
The first forbidden $\beta$-decay transitions can be defined in
terms of multipole operator. For the $2^{-}\rightarrow 0^{+}$
transitions these are
\begin{eqnarray}
M^{U1F}_{\beta^{\mp}}=
M^{\mp}(j_{A},\kappa=1,\lambda=2,\mu)=g_{A}\sum_{k=1}^{A}t_{\mp}(k)r_{k}\{Y_{1}(r_{k})\sigma(k)\}_{2\mu}.
\end{eqnarray}
Here $M^{\mp}(j_{A},\kappa=1,\lambda=2,\mu)$ is the non-relativistic
unique first forbidden $\beta$-decay multipole operator
\cite{Boh69}. All symbols have their usual meanings. The transitions
probability $B(I_{i}\longrightarrow I_{f}, \beta^{\mp})$ is
described by the reduced matrix element of the multipole operator
(Eq.(7)). Thus, we may write
\begin{eqnarray}
B(I_{i}\longrightarrow I_{f},
\beta^{\mp})=\frac{1}{2I_{i}+1}|<I_{f}\|M^{\mp}(j_{A},\kappa=1,\lambda=2)\|I_{i}>|^{2}.
\end{eqnarray}
The reduced matrix elements $\langle 2^{-}_{i}
\|M_{\beta^{\mp}}\|0^{+}\rangle$ within framework of the pn-QRPA(WS)
method are given as
\begin{eqnarray}
<2^{-}_{i}|M_{\beta^{-}}(\mu')|0^{+}>=<0^{+}|[Q_{i}(\mu),M_{\beta^{-}}(\mu')]|2>=\nonumber
\end{eqnarray}
\begin{eqnarray}
\sum_{j_{p}j_{n}}\delta_{\mu\mu'}\{b_{j_{p}j_{n}}\psi^{i}_{j_{p}j_{n}}(\mu)+\bar{b}_{j_{p}j_{n}}\varphi^{i}_{j_{p}j_{n}}(\mu)\},\nonumber
\end{eqnarray}
\begin{eqnarray}
<2^{-}_{i}|M_{\beta^{+}}(\mu')|0^{+}>=<0^{+}|[Q_{i}(\mu),M_{\beta^{+}}(\mu')]|2>=\nonumber
\end{eqnarray}
\begin{eqnarray}
\sum_{j_{p}j_{n}}\delta_{\mu\mu'}\{\bar{b}_{j_{p}j_{n}}\psi^{i}_{j_{p}j_{n}}(\mu)+b_{j_{p}j_{n}}\varphi^{i}_{j_{p}j_{n}}(\mu)\}.\nonumber
\end{eqnarray}
Transitions with $\lambda=n+1$ are referred to as unique first
forbidden transitions \cite{Boh69}, and the $ft$ values are
expressed as
\begin{eqnarray}
(ft)_{\beta^{\mp}}=\frac{D}{(g_{A}/g_{V})^{2}4\pi
B(I_{i}\longrightarrow I_{f},
\beta^{\mp})}~\frac{(2n+1)!!}{[(n+1)!]^{2}n!},\nonumber
\end{eqnarray}
where
\begin{eqnarray}
D=\frac{2\pi^{3}\hbar^{2}ln2}{g_{V}^{2}m_{e}^{5}c^{4}}=6250sec,~~~\frac{g_{A}}{g_{V}}=-1.254.\nonumber
\end{eqnarray}

\subsection{Allowed GT, U1F transitions and stellar $\beta$-decay rates in the pn-QRPA(N) model}
In the pn-QRPA(N) model \cite{Mut92}, proton-neutron residual
interactions take place as $ph$ (characterized by interaction
constant $\chi$) and $pp$ (characterized by interaction constant
$\kappa$) interactions. As in the pn-QRPA(WS) model,  we also use
separable interaction here. The QRPA matrix equation converts to an
algebraic equation of fourth order by using these separable GT
forces, that is much simpler to solve as compared to full
diagonalization of the non-Hermitian matrix having very large
dimensions \cite{Hom96, Mut92}. A quasiparticle basis was first
constructed (described by a Bogoliubov transformation) having a
pairing interaction, and then the RPA equation having schematic
separable GT residual interaction was solved. The single particle
energies were determined employing a deformed Nilsson oscillator
potential with quadratic deformation. The pairing correlation was
taken into account in the BCS approximation utilizing constant
pairing forces. The BCS calculations were performed on the deformed
Nilsson basis for neutrons and protons independently. The detailed
formalism for calculation of allowed weak rates in astrophysical
environment using the pn-QRPA(N) model  can be seen in Refs.
\cite{Nab99, Nab04}.

For the calculation of the U1F weak-decay rates, nuclear matrix
elements of the separable forces which occur in RPA equation are
given by

\begin{equation}
V^{ph}_{pn,p^{\prime}n^{\prime}} = +2\chi
f_{pn}(\mu)f_{p^{\prime}n^{\prime}}(\mu),
\end{equation}

\begin{equation}
V^{pp}_{pn,p^{\prime}n^{\prime}} = -2\kappa
f_{pn}(\mu)f_{p^{\prime}n^{\prime}}(\mu),
\end{equation}

where
\begin{equation}
f_{pn}(\mu)=<j_{p}m_{p}|t_{-}r[\sigma Y_{1}]_{2\mu}|j_{n}m_{n}>,
\end{equation}
is a single-particle U1F transition amplitude (the symbols have
their usual meaning). It should be noted that $\mu$ can take the
values of 0,$\pm1$, and $\pm2$, and the neutron and proton states
have opposite parities, (however for allowed weak rates $\mu$ only
takes the values $0$ and $\pm1$) \cite{Hom96}. The $ph$ interaction
constant $\chi$ was considered to be $4.2/A$ MeV and $56.16/A$ MeV
fm$^{-2}$, for allowed and  U1F transitions, respectively, following
a $1/A$ dependence \cite{Hom96}. The functional form of $\chi$ is
the same as that used in the recent pn-QRPA(N) model calculation of
\cite{Nab16}. For describing the $\beta^{+}$ (and
$\beta$$\beta$)-decay the $pp$ force is important \cite{Eng88,
Mut89a}. We therefore take into account the $pp$ force, the $pp$
interaction constant $\kappa$ was taken to be 0.01. These values of
$\chi$ and $\kappa$ well reproduced the experimental half-lives
values.

Nuclear deformation was calculated by using
\begin{equation}
\delta = \frac{125(Q_{2})}{1.44 (Z) (A)^{2/3}},
\end{equation}
where $Z$ and $A$ represents the charge and mass numbers,
respectively and the electric quadrupole moment $Q_{2}$ values are
taken from Ref. \cite{Mol81}. Q-values were taken from Audi
\textit{et al}. \cite{Aud12}.

The stellar U1F weak rates from the  $\mathit{i}$th parent state to
the  $\mathit{j}$th daughter state of the nucleus are given by

\begin{equation}
\lambda_{ij} = \frac{m_{e}^{5}c^{4}}{2\pi^{3}\hbar^{7}}\sum_{\Delta
J^{\pi}}g^{2}f(\Delta J^{\pi};ij)B(\Delta J^{\pi};ij),
\end{equation}
where $f(\Delta J^{\pi};ij)$ and $ B(\Delta J^{\pi};ij)$ are the
integrated Fermi function and $\beta$-decay reduced transition
probability, correspondingly, for the transition $i \rightarrow j$
which induces a spin-parity change $\Delta J^{\pi}$ and $g$
represent the weak coupling constant which took the value of $g_{A}$
or $g_{V}$ conforming to whether the $\Delta J^{\pi}$ transition is
linked with the axial-vector or vector weak-interaction. The
phase-space factors $f(\Delta J^{\pi};ij)$ are taken as integrals
over the lepton distribution functions and therefore are sensitive
to the temperature as well as density inside the stellar medium. The
$B(\Delta J^{\pi};ij)$ are linked with the U1F weak interaction
matrix elements discussed earlier.

For the U1F transitions the phase space integrals are given by

\begin{eqnarray}
f = \int_{1}^{w_{m}} w \sqrt{w^{2}-1}
(w_{m}-w)^{2}[(w_{m}-w)^{2}F_{1}(Z,w) \nonumber\\
+ (w^{2}-1)F_{2}(Z,w)] (1-G_{-}) dw,
\end{eqnarray}
where $w$ show the total kinetic energy (K.E) of the electron having
its rest mass and $w_{m}$ shows the total $\beta$-decay energy ($
w_{m} = m_{p}-m_{d}+E_{i}-E_{j}$, where $m_{p}$ and $E_{i}$ are the
mass and excitation energies of the parent nucleus, and $m_{d}$ and
$E_{j}$ of the daughter nucleus, correspondingly). $G_{-}$ are the
electron distribution functions. Assume that the electrons are not
in the bound state, these are the Fermi-Dirac distribution
functions,
\begin{equation}
G_{-} = [exp(\frac{E-E_{f}}{kT})+1]^{-1}.
\end{equation}
Here $E=(w-1)$ is the K.E of the electrons, $E_{f}$ is the Fermi
energy of the electrons, $T$ is the temperature, and $k$ is the
Boltzmann constant.

The Fermi functions, $F_{1}(\pm Z,w)$ and $F_{2}(\pm Z,w)$ appearing
in Eq.~(14) were computed according to the procedure used by
\cite{Gov71}. The electrons  number density related with nuclei and
protons is $\rho Y_{e} N_{A}$, where $\rho$ represent the baryon
density, $Y_{e}$ is the ratio of electron number to the baryon
number, and $N_{A}$ is the Avogadro's number.
\begin{equation}
\rho Y_{e} = \frac{1}{\pi^{2}N_{A}}(\frac {m_{e}c}{\hbar})^{3}
\int_{0}^{\infty} (G_{-}-G_{+}) p^{2}dp,
\end{equation}
here $p=(w^{2}-1)^{1/2}$ show the positron or electron momentum,
 $G_{+}$ is the positron distribution
functions specified by
\begin{equation} G_{+} =\left[\exp
\left(\frac{E+2+E_{f} }{kT}\right)+1\right]^{-1}.
\end{equation}
Eq.~(16) is used for the iterative calculation of Fermi energies at
particular values of $T$ and $\rho Y_{e}$.

As a result of very high temperatures prevailing in the stellar
medium, there is a finite probability of occupation of parent
excited states in the core of massive stars.  There is hence a
finite contribution to weak-decay rates from these excited states.
On the assumption of thermal equilibrium the occupation probability
of a state $i$ is estimated as

\begin{equation}
 P_{i} = \frac {exp(-E_{i}/kT)}{\sum_{i=1}exp(-E_{i}/kT)},
 \end{equation}
where $E_{i}$ represent the excitation energy of the state $i$.
Finally, for any weak processes the rate per unit time per nucleus
is  obtained as
\begin{equation} \lambda = \sum_{ij}P_{i}
\lambda_{ij}.
\end{equation}
The summation is carried out on all initial and final states until
satisfactory convergence in the rates calculation is obtained. It is
noted that we use a huge model space (up to 7 major oscillator
shells) in the pn-QRPA(N) model due to which convergence is readily
achieved in our weak rates calculation for excitation energies well
in excess of 10 MeV (for parent as well as daughter states). For
further details of the pn-QRPA(N) formalism we refer to \cite{Nab16}
which we do not repeat for space consideration.

\section{Results and Discussions}
As mentioned earlier we quench all our charge-changing transitions
by a quenching factor of $0.7$ in both pn-QRPA(WS) and pn-QRPA(N)
models. All the pn-QRPA calculations, which we shall discuss in this
section, were performed for a total of  26 (22 odd-odd and 4
even-even) nuclei with mass range $70\leq A \leq 214$. We begin the
proceedings by comparing our calculated reduced matrix elements (for
$\Delta J=2$ transitions), within the pn-QRPA(WS) and pn-QRPA(N)
models, with extracted reduced matrix elements from experimental
$logft$ values \cite{Led78}. The results are shown in Table~1. It is
to be noted that there are 29 entries in Table~1. This is because
for $^{84}$Rb, $^{102}$Rh and $^{122}$Sb, we calculate U1F
transitions in both $\beta$-decay and electron capture directions.
Our matrix elements are also compared with the calculated
non-relativistic matrix elements taken from \cite{Civ86} and shown
as RPA in seventh column of Table~1. The last two columns show our
calculated matrix elements.  The root mean square deviations for
pn-QRPA(WS) and pn-QRPA(N) calculated results are 0.1876 and 0.0966,
respectively. It is clear from Table~1 that the calculated reduced
matrix elements of pn-QRPA(N) model are in better agreement with the
experimental results. The pn-QRPA(WS) model calculated matrix
elements are in better agreement with the experimental values as
compared to the correlated RPA values.

The calculated $\log ft$ values of pn-QRPA(N) and pn-QRPA(WS) models
are compared with the experimental values as well as with the
theoretical calculation of \cite{Civ86} in Fig.~1. The figure shows
that both pn-QRPA model results are in better agreement with the
measured $\log ft$ values. The calculated $logft$ values by
\cite{Civ86} are up to two orders of magnitude smaller than the
measured data. The assumption in \cite{Civ86}, that the relativistic
$\beta$-moment  is proportional to the matrix element of
non-relativistic $\beta$-moment, is the major source of the orders
of magnitude disagreement with the experimental data.

The collective state strength distributions, as a function of energy
for some $2^{-}\longrightarrow 0^{+}$ transitions, calculated using
the pn-QRPA(WS) model are compared with the RPA results \cite{Civ86}
in Fig.~2. The main contributions to the strength are situated at
energies of the order (21--25) MeV and shows the position of the
giant FF resonance (FFR) in our calculations with $I^{\pi} = 2^{-}$.
This value is incidentally very close to the measured values
determined by Horen et al. \cite{Hor1, Hor2}. For these transitions,
the corresponding average peak energy for the $2^{-}$ giant FFR has
been reported around (19--22) MeV \cite{Civ86}. Our calculated
strength distributions are in reasonable agreement with the RPA
results.

The pn-QRPA(N) model takes into account the nuclear deformation.
This results in a  fragmentation of $\beta$-decay strength
distribution \cite{Sta90, Hir93}. In order to improve the
reliability of our calculated weak rates, experimental data (XUNDL)
were included in our pn-QRPA(N) model calculation wherever possible.
The calculated excitation energies were substituted with measured
levels when they were within 0.5 MeV of each other. Missing measured
levels were inserted  (along with their log$ft$ values) wherever
possible. The UIF transitions for $\beta$-decay of $^{82}$Br,
$^{94}$Y, $^{136}$I and $^{204}$Au using the pn-QRPA(N) model are
shown in Fig.~3. Here the abscissa represent the daughter excitation
energy in units of MeV. Similarly Fig.~4 shows the calculated  UIF
transitions for electron capture (EC) of $^{72}$As and $^{198}$Tl.
All U1F transitions are shown in units of $fm^{2}$ up to 30 MeV in
daughter nuclei. The calculated U1F transition strength for
$^{72}$As (EC direction) and $^{82}$Br ($\beta$-decay direction)
using the pn-QRPA(N) model in spherical limit (setting deformation
parameter $\beta$ equal to zero)  are shown in Fig.~5. It is clear
from Fig.~5 that deformation leads to much fragmented data. The
calculated UIF strength distribution in $\beta$-decay and EC
direction, using the pn-QRPA(WS) model are shown in Fig.~6 and
Fig.~7, respectively. As discussed earlier the pn-QRPA(WS) model
considered only spherical nuclei, in which the Woods-Saxon potential
basis was used. The calculated strength distributions for deformed
nuclei are more fragmented than those calculated without the
deformation.

For the first time we calculate the  allowed GT and U1F weak
interaction rates for all heavy nuclei (shown in Table~1) using the
pn-QRPA(N) model in stellar environment. These include electron and
positron emission rates as well as electron and positron capture
rates. The weak-rate calculations were performed at stellar density
range (10--10$^{11})$g/cm $^{3}$ and at stellar temperature in the
range 0.01 $\leq$ T$_{9}$ $\leq$ 30 (T$_{9}$ represents the stellar
temperature in units of 10$^{9}$ K). We  plan to calculate the
beta-delayed neutron emission rates for heavier nuclei in near
future. Tables~2 and 3 show the total allowed rates in $\beta$-decay
direction (sum of positron capture (PC) and $\beta^{-}$-decay rates)
and the percentage contribution of $\beta^{-}$-decay to total rate
for 20 heavy nuclei. Table~4, on the other hand, shows the
calculation of weak rates along the electron capture direction (sum
of electron capture (EC) and $\beta^{+}$-decay rates) for allowed
transitions for  9 heavy nuclei. The corresponding  U1F weak rates
calculation are shown in Tables (5--7). The rates are shown at
selected densities (10$^{3}$, 10$^{7}$ and 10$^{11}$) g/cm$^{3}$ and
at temperatures (1.5, 5, 10 and 30) GK. Tables 2 and 3 show that the
sum of PC and $\beta^{-}$-decay rates increases with increasing
temperature, but decreases with increase in density. The growth of
the stellar density suppresses the rates due to the availability of
smaller phase space, whereas increase in temperature weakens the
effect of Pauli blocking and consequently enhances the contribution
of the GT$_{-}$ transitions from excited levels of parent nucleus.
As the temperature increases the PC rates contribution to the total
rate also increases. This makes sense as positrons are produced only
at high enough temperatures ($kT >$ 1 MeV). At a fixed temperature
the percentage contribution of $\beta^{-}$-decay rates increases
with increasing stellar density. For high stellar temperatures (T$_9
\ge 5$) our calculation shows that PC rates should not be neglected
and contribute significantly to the total rates specially for low
stellar densities. Regarding U1F transitions, at low stellar
temperatures, the total rates are commanded by $\beta^{-}$-decay and
at high temperatures by PC rates (see Tables 5 and 6).

Tables~4 and 7 depict that the sum of EC and $\beta^{+}$-decay rates
increases with increasing temperature and density. The weak rates go
up as the temperature increases because more excited levels
contribute to the total rate as the stellar temperature rises. Also
as the density and temperature of the core increase, the Fermi
energy of electrons increases due to which enhancement of EC rates
occurs. Consequently the contribution of EC rates to total weak
rates becomes very large at high density and temperature regions. It
is seen from these tables that at T$_9$=30 the contribution of
$\beta^{+}$-decay to total rates can safely be neglected.

Figures 8 and 9 show the contribution of allowed and U1F rates to
total transition probabilities. The bar graphs are shown at stellar
density 10$^{7}$ g/cm$^{3}$ and at temperatures 5 $\times$ 10$^{9}$
K. For all nuclei large contribution of U1F rates to total rates is
seen at low and higher temperature region.  At higher temperature
(T$_9$=30) the contribution of U1F rates increases further both at
low and high density regions. Our calculation shows that for some
nuclei, forbidden transitions have big contributions to total rates.
The significant contribution of U1F transition to total
$\beta$-transition probability, for many nuclei, was also calculated
by Homma \emph{et al.} (see Fig. 9 and Table IX) \cite{Hom96}. The
weak rates reported here has contributions only from U1F
transitions. It is desirable to study the effects of non-unique
forbidden transitions to the calculated $\beta$-decay half-lives and
which would be taken up as a future project.

\section{Conclusions}
Generally for heavier nuclei (as the neutron number increases) the
contribution of U1F strength becomes more and more important. The
$2^-\longleftrightarrow 0^+$ U1F weak decay strength for odd-odd and
even-even nuclei in the mass range $70\leq A \leq214$ was calculated
using two different pn-QRPA models. The pn-QRPA(WS) model considered
only spherical nuclei while deformed nuclei were considered within
the pn-QRPA(N) model. It was concluded that the U1F transitions have
significant contribution to the beta decay half-lives. The $ft$
values and reduced matrix elements for the $\Delta J=2$ transitions,
in the mass region $70\leq A \leq214$, were calculated using both
pn-QRPA models. The pn-QRPA calculated results agreed well with the
experimental results and proved to be a considerable improvement
over previous RPA calculations. A  major concentration  of  the
correlated strength  was  found  at energies of  the  order  of the
giant FFR. The  energy  of  the associated giant  FFR  was found  to
be in good agreement  with the experimental  and model estimates.
For stellar applications we calculated the allowed and U1F weak
rates using the pn-QRPA(N) model. It was concluded that electron and
positron emission rates can safely be neglected at high stellar
temperatures. Supernova simulators are urged to test run our
reported rates for probable interesting outcomes. The ASCII files of
all calculated terrestrial and stellar rates may be requested from
the authors.

\section*{Acknowledgements}
J.-U. Nabi wishes to acknowledge the support provided by the Higher
Education Commission (Pakistan) through the HEC Project No. 20-3099.

\clearpage \onecolumn
\begin{table}
\tiny \caption{Experimental and theoretical results for $\Delta J =
2$ first-forbidden $\beta$-decay transitions. The $\Delta J=2\
[I_{i}\ (I_{f})\ =2^{-}\ (0^{+})]$ transition which have been
considered are shown in the first two columns. The experimental log
\emph{ft} values \cite{Led78} and the extracted reduced matrix
elements are shown in columns three and four. The correlated RPA
values of the reduced matrix elements \cite{Civ86} are shown in
column seven while last two columns show the reduced matrix elements
of pn-QRPA(WS) and pn-QRPA(N) models, respectively.} \label{Table 1}
\begin{center}
\scalebox{0.7}

\begin{tabular}
{ccccccccccc} \hline \\$N$ & \multicolumn{2}{c}{Transition} &
$log\textit{ft}_{ex}$ & $<I_{f}\Vert M^{U1F}\Vert I_{i}>_{ex}$ &
\multicolumn{2}{c}{s-p transition} & &
\multicolumn{3}{c}{$<I_{f}\Vert M^{U1F}\Vert I_{i}>_{th}[$fm$]$
}\\\\\cline{2-3} \cline{6-7} \cline{9-11}\\
 & initial & final &  & [fm] &
proton & neutron & & RPA & pn-QRPA(WS) & pn-QRPA(N)\\
\\
\hline
1  & 72As(2$^{-}$)   & 72Ge(0$^{+}$)  & 9.80  & 0.2029  & 1f$_{5/2}$  & 1g$_{9/2}$ &  & 0.4254   & 0.1720  & 0.2355 \\
2  & 82Br(2$^{-}$)   & 82Kr(0$^{+}$)  & 8.90  & 0.5988  & 1f$_{5/2}$  & 1g$_{9/2}$  & & 1.9297   & 0.5287  & 0.4546 \\
3  & 84Br(2$^{-}$)   & 84Kr(0$^{+}$)  & 9.50  & 0.3106  & 1f$_{5/2}$  & 1g$_{9/2}$  & & 1.3480   & 0.4404  & 0.2987 \\
4  & 86Br(2$^{-}$)   & 86Kr(0$^{+}$)  & 9.07  & 0.3001  & 2p$_{3/2}$  & 2d$_{5/2}$  & & 0.9642   & 0.4086  & 0.2864 \\
5  & 84Rb(2$^{-}$)   & 84Sr(0$^{+}$)  & 9.40  & 0.3445  & 1f$_{5/2}$  & 1g$_{9/2}$  & & 2.0324   & 0.5478  & 0.3904 \\
6 & 84Rb(2$^{-}$)   & 84Kr(0$^{+}$)  & 9.50  & 0.3036  & 1f$_{5/2}$  & 1g$_{9/2}$  & & 1.5500   & 0.3501  & 0.1651 \\
7  & 86Rb(2$^{-}$)   & 86Sr(0$^{+}$)  & 9.40  & 0.3367  & 1f$_{5/2}$  & 1g$_{9/2}$  & & 1.4530   & 0.4552  & 0.5019 \\
8  & 88Rb(2$^{-}$)   & 88Sr(0$^{+}$)  & 9.20  & 0.4239  & 2p$_{1/2}$  & 2d$_{5/2}$  & & 0.5150   & 0.4238  & 0.5147 \\
9  & 90Y(2$^{-}$)    & 90Zr(0$^{+}$)  & 9.20  & 0.4239  & 2p$_{1/2}$  & 2d$_{5/2}$  & & 1.2113   & 0.5330  & 0.4555 \\
10  & 92Y(2$^{-}$)    & 92Zr(0$^{+}$)  & 9.20  & 0.3938  & 2p$_{1/2}$  & 2d$_{5/2}$  & & 1.1722   & 0.4365  & 0.5804 \\
11 & 94Y(2$^{-}$)    & 94Zr(0$^{+}$)  & 9.30  & 0.3778  & 2p$_{1/2}$  & 2d$_{5/2}$  & & 1.0989   & 0.5136  & 0.1895 \\
12 & 90Sr(0$^{+}$)   & 90Y(2$^{-}$)   & 9.40  & 0.1506  & 2p$_{1/2}$  & 2d$_{5/2}$  & & 0.2395   & 0.1194  & 0.1454 \\
13 & 92Sr(0$^{+}$)   & 92Y(2$^{-}$)   & 8.90  & 0.5995  & 2p$_{1/2}$  & 2d$_{5/2}$  & & 0.3422   & 0.1788  & 0.5154 \\
14 & 88Kr(0$^{+}$)   & 88Rb(2$^{-}$)  & 9.30  & 0.1690  & 2p$_{3/2}$  & 2d$_{5/2}$  &  &0.0937   & 0.3365  & 0.1604 \\
15 & 102Rh(2$^{-}$)  & 102Ru(0$^{+}$) & 9.70  & 0.5588  & 2p$_{1/2}$  & 2d$_{5/2}$  & & 0.4271   & 0.2111  & 0.6788 \\
16 & 102Rh(2$^{-}$)  & 102Pd(0$^{+}$) & 9.70  & 0.2439  & 2p$_{1/2}$  & 2d$_{5/2}$  & & 1.0570   & 0.3438  & 0.2388 \\
17 & 120I(2$^{-}$)   & 120Te(0$^{+}$) & 9.30  & 0.3778  & 1g$_{7/2}$  & 1h1$_{1/2}$ & & 1.8610   & 0.3720  & 0.2534 \\
18 & 124I(2$^{-}$)   & 124Te(0$^{+}$) & 9.30  & 0.3778  & 1g$_{7/2}$  & 1h1$_{1/2}$ & & 2.7326   & 0.6150  & 0.3955 \\
19 & 126I(2$^{-}$)   & 126Te(0$^{+}$) & 9.13  & 0.4594  & 1g$_{7/2}$  & 1h1$_{1/2}$ & & 3.1254   & 0.7230  & 0.4040 \\
20 & 136I(2$^{-}$)   & 136Xe(0$^{+}$) & 8.63  & 0.1487  & 1g$_{7/2}$  & 1h1$_{1/2}$ & & 2.3621   & 0.4520  & 0.1313 \\
21 & 122Sb(2$^{-}$)  & 122Sn(0$^{+}$) & 8.90  & 0.5988  & 1g$_{7/2}$  & 1h1$_{1/2}$ & & 3.2304   & 0.6453  & 0.6879 \\
22 & 122Sb(2$^{-}$)  & 122Te(0$^{+}$) & 8.60  & 0.2674  & 1g$_{7/2}$  & 1h1$_{1/2}$ & & 1.7336   & 0.4680  & 0.2894 \\
23 & 132La(2$^{-}$)  & 132Ba(0$^{+}$) & 9.50  & 0.3001  & 1g$_{7/2}$  & 1h1$_{1/2}$ & & 2.0986   & 0.4540  & 0.1961 \\
24 & 140Ba(0$^{+}$)  & 140La(2$^{-}$) & 8.82  & 0.2936  & 1g$_{7/2}$  & 1h$_{9/2}$  & & 0.1944   & 0.4060  & 0.3069 \\
25 & 142Pr(2$^{-}$)  & 142Nd(0$^{+}$) & 8.90  & 0.5987  & 1g$_{7/2}$  & 1h$_{9/2}$  & & 1.4403   & 0.4600  & 0.5743 \\
26 & 198Au(2$^{-}$)  & 198Hg(0$^{+}$) & 11.20 & 0.0119  & 3s$_{1/2}$  & 3p$_{7/2}$  & & 0.0777   & 0.1880  & 0.0151 \\
27 & 204Au(2$^{-}$)  & 204Hg(0$^{+}$) & 8.50  & 0.9489  & 2d$_{7/2}$  & 3p$_{1/2}$  & & 0.3660   & 0.6370  & 0.7151 \\
28 & 198Tl(2$^{-}$)  & 198Hg(0$^{+}$) & 9.00  & 0.1893  & 3s$_{1/2}$  & 3p$_{7/2}$  & & 1.8409   & 0.4730  & 0.2585 \\
29 & 204Tl(2$^{-}$)  & 204Pb(0$^{+}$) & 10.10 & 0.1513  & 1h$_{9/2}$  & 1i1$_{7/2}$ & & 0.8894   & 0.3370  & 0.1413 \\
\hline
\end{tabular}
\end{center}
\end{table}

\onecolumn
\begin{table}
\scriptsize \caption{Allowed rates for different selected densities
and temperatures. The second column gives stellar densities
($\rho$Y$_{e}$) having units of g/cm$^{3}$, where $\rho$ is the
baryon density and Y$_{e}$ is the ratio of the electron number to
the baryon number. Temperatures (T$_{9}$) are given in units of
10$^{9}$ K. $\lambda_{total}$ shows sum of $\beta^{-}$ and positron
capture (PC) rates.}\label{Table 2}

\begin{tabular}{c|c|cccc|cccc}

Nucleus & $\rho$$\it Y_{e}$ & \multicolumn{4}{c|}{$\lambda_{total} (s^{-1})$} & \multicolumn{3}{c}{Percentage contribution of $\beta^{-}$-decay} \\
\cline{3-10} & &T$_{9}$=1.5 & T$_{9}$=5 & T$_{9}$=10 & T$_{9}$=30 &T$_{9}$=1.5 & T$_{9}$=5 & T$_{9}$=10 & T$_{9}$=30\\
\hline
         & $10^{3}$ & 7.22E-07 & 1.07E-03 & 8.93E-02 & 4.35E+02 & 6.15E+01 & 3.07E+01 & 1.55E+01 & 5.17E-02 \\
$^{82}$Br & $10^{7}$ & 7.77E-08 & 4.30E-04 & 7.38E-02 & 4.32E+02 & 1.00E+02 & 6.52E+01 & 1.82E+01 & 5.20E-02 \\
      & $10^{11}$ & 1.55E-82 & 5.79E-26 & 3.05E-13 & 5.98E-02 & 1.00E+02 & 9.90E+01 & 7.53E+01 & 1.70E-01 \\
      \hline
      & $10^{3}$ & 6.78E-05 & 5.36E-03 & 2.43E-01 & 5.77E+02 & 9.77E+01 & 6.20E+01 & 3.59E+01 & 2.77E-01 \\
$^{84}$Br & $10^{7}$ & 1.02E-05 & 3.31E-03 & 2.10E-01 & 5.72E+02 & 1.00E+02 & 8.76E+01 & 4.07E+01 & 2.78E-01 \\
      & $10^{11}$ & 4.65E-77 & 2.62E-24 & 2.96E-12 & 8.03E-02 & 1.00E+02 & 9.99E+01 & 9.48E+01 & 1.04E+00 \\
\hline
      & $10^{3}$ & 1.77E-02 & 6.86E-02 & 7.25E-01 & 3.33E+02 & 1.00E+02 & 9.36E+01 & 6.64E+01 & 7.09E-01 \\
$^{86}$Br & $10^{7}$ & 1.25E-02 & 6.04E-02 & 6.71E-01 & 3.30E+02 & 1.00E+02 & 9.85E+01 & 7.09E+01 & 7.14E-01 \\
      & $10^{11}$ & 2.02E-67 & 1.58E-21 & 6.31E-11 & 4.75E-02 & 1.00E+02 & 1.00E+02 & 9.96E+01 & 3.53E+00 \\
\hline
      & $10^{3}$ & 4.57E-05 & 3.91E-03 & 4.60E-01 & 1.65E+02 & 9.62E+01 & 3.88E+01 & 6.54E+01 & 1.99E+00 \\
$^{88}$Kr & $10^{7}$ & 2.37E-05 & 1.92E-03 & 4.25E-01 & 1.64E+02 & 1.00E+02 & 7.50E+01 & 7.00E+01 & 2.00E+00 \\
      & $10^{11}$ & 4.17E-74 & 1.14E-22 & 5.29E-11 & 2.50E-02 & 1.00E+02 & 1.00E+02 & 9.97E+01 & 1.00E+01 \\
\hline
      & $10^{3}$ & 7.76E-12 & 5.15E-05 & 2.22E-02 & 2.90E+02 & 2.57E-01 & 9.15E+00 & 3.17E+00 & 6.68E-03 \\
$^{84}$Rb & $10^{7}$ & 3.07E-15 & 1.30E-05 & 1.79E-02 & 2.88E+02 & 6.99E+01 & 2.86E+01 & 3.77E+00 & 6.71E-03 \\
      & $10^{11}$ & 5.17E-90 & 2.22E-28 & 2.69E-14 & 3.99E-02 & 9.94E+01 & 8.32E+01 & 2.14E+01 & 1.88E-02 \\
\hline
      & $10^{3}$ & 6.62E-10 & 1.52E-04 & 3.87E-02 & 3.73E+02 & 2.95E+01 & 2.34E+01 & 7.52E+00 & 1.20E-02 \\
$^{86}$Rb & $10^{7}$ & 1.46E-11 & 5.26E-05 & 3.14E-02 & 3.70E+02 & 9.96E+01 & 5.61E+01 & 8.90E+00 & 1.21E-02 \\
      & $10^{11}$ & 1.04E-86 & 2.17E-27 & 6.13E-14 & 5.12E-02 & 1.00E+02 & 9.57E+01 & 4.23E+01 & 3.47E-02 \\
\hline
      & $10^{3}$ & 9.06E-04 & 1.48E-02 & 3.64E-01 & 6.52E+02 & 9.99E+01 & 9.06E+01 & 5.22E+01 & 2.47E-01 \\
$^{88}$Rb & $10^{7}$ & 4.09E-04 & 1.25E-02 & 3.26E-01 & 6.47E+02 & 1.00E+02 & 9.77E+01 & 5.72E+01 & 2.48E-01 \\
      & $10^{11}$ & 1.19E-74 & 1.87E-23 & 6.48E-12 & 9.06E-02 & 1.00E+02 & 1.00E+02 & 9.73E+01 & 9.33E-01 \\
\hline
      & $10^{3}$ & 1.84E-09 & 5.92E-04 & 1.86E-01 & 1.85E+02 & 5.13E-02 & 6.41E+01 & 5.27E+01 & 6.64E-01 \\
$^{90}$Sr & $10^{7}$ & 7.77E-13 & 3.97E-04 & 1.67E-01 & 1.83E+02 & 7.17E+01 & 8.94E+01 & 5.78E+01 & 6.68E-01 \\
      & $10^{11}$ & 2.44E-81 & 1.03E-24 & 4.08E-12 & 2.60E-02 & 1.00E+02 & 1.00E+02 & 9.79E+01 & 2.60E+00 \\
\hline
      & $10^{3}$ & 5.12E-05 & 2.11E-03 & 1.98E-01 & 8.05E+01 & 9.99E+01 & 8.82E+01 & 7.31E+01 & 1.09E+00 \\
$^{92}$Sr & $10^{7}$ & 1.35E-05 & 1.82E-03 & 1.85E-01 & 7.98E+01 & 1.00E+02 & 9.73E+01 & 7.70E+01 & 1.10E+00 \\
      & $10^{11}$ & 8.32E-76 & 1.72E-23 & 1.01E-11 & 1.15E-02 & 1.00E+02 & 1.00E+02 & 9.95E+01 & 4.81E+00 \\
\hline
      & $10^{3}$ & 6.81E-09 & 3.20E-04 & 3.89E-02 & 3.78E+02 & 7.74E+00 & 5.88E+01 & 8.74E+00 & 7.06E-03 \\
$^{90}$Y & $10^{7}$ & 1.69E-10 & 1.75E-04 & 3.15E-02 & 3.75E+02 & 9.96E+01 & 8.48E+01 & 1.02E+01 & 7.11E-03 \\
      & $10^{11}$ & 3.63E-85 & 4.16E-27 & 5.49E-14 & 5.19E-02 & 1.00E+02 & 9.74E+01 & 3.58E+01 & 1.86E-02 \\
\hline
\end{tabular}
\end{table}

\onecolumn
\begin{table}
\scriptsize \caption{Same as Table~2.}\label{Table 3}

\begin{tabular}{c|c|cccc|cccc}

Nucleus & $\rho$$\it Y_{e}$ & \multicolumn{4}{c|}{$\lambda_{total} (s^{-1})$} & \multicolumn{3}{c}{Percentage contribution of $\beta^{-}$-decay} \\
\cline{3-10} & &T$_{9}$=1.5 & T$_{9}$=5 & T$_{9}$=10 & T$_{9}$=30 &T$_{9}$=1.5 & T$_{9}$=5 & T$_{9}$=10 & T$_{9}$=30\\
\hline
& $10^{3}$ & 1.03E-07 & 5.42E-03 & 1.60E-01 & 4.98E+02 & 9.63E+01 & 9.39E+01 & 3.82E+01 & 6.95E-02 \\
$^{92}$Y & $10^{7}$ & 2.79E-08 & 4.50E-03 & 1.38E-01 & 4.95E+02 & 1.00E+02 & 9.85E+01 & 4.27E+01 & 6.98E-02 \\
      & $10^{11}$ & 6.47E-80 & 4.47E-25 & 6.97E-13 & 6.87E-02 & 1.00E+02 & 9.99E+01 & 8.58E+01 & 2.04E-01 \\
\hline
      & $10^{3}$ & 8.02E-04 & 1.22E-02 & 7.58E-02 & 1.24E+02 & 1.00E+02 & 9.60E+01 & 5.33E+01 & 1.84E-01 \\
$^{94}$Y & $10^{7}$ & 5.06E-04 & 1.05E-02 & 6.76E-02 & 1.23E+02 & 1.00E+02 & 9.91E+01 & 5.80E+01 & 1.85E-01 \\
      & $10^{11}$ & 2.36E-75 & 2.35E-24 & 6.80E-13 & 1.72E-02 & 1.00E+02 & 1.00E+02 & 9.47E+01 & 5.99E-01 \\
\hline
      & $10^{3}$ & 2.11E-09 & 1.44E-04 & 2.63E-01 & 1.11E+03 & 3.04E+00 & 2.63E+00 & 5.52E-01 & 5.26E-04 \\
$^{102}$Rh & $10^{7}$ & 1.36E-12 & 3.10E-05 & 3.29E-01 & 1.12E+03 & 8.21E+01 & 9.22E+00 & 6.61E-01 & 5.27E-04 \\
      & $10^{11}$ & 4.99E-88 & 2.57E-28 & 2.37E+05 & 1.05E+06 & 9.83E+01 & 5.58E+01 & 4.73E+00 & 1.55E-03 \\
\hline
      & $10^{3}$ & 2.00E-06 & 1.95E-03 & 1.12E-01 & 5.22E+02 & 2.17E+01 & 2.50E+00 & 3.46E-01 & 3.07E-04 \\
$^{122}$Sb & $10^{7}$ & 8.20E-09 & 4.17E-04 & 8.97E-02 & 5.19E+02 & 9.77E+01 & 8.27E+00 & 4.11E-01 & 3.08E-04 \\
      & $10^{11}$ & 2.02E-84 & 2.82E-27 & 1.15E-13 & 7.16E-02 & 9.97E+01 & 4.52E+01 & 2.47E+00 & 8.41E-04 \\
\hline
      & $10^{3}$ & 1.85E-02 & 5.72E-01 & 2.29E+00 & 1.19E+03 & 1.00E+02 & 9.85E+01 & 6.56E+01 & 2.45E-01 \\
$^{136}$I & $10^{7}$ & 1.30E-02 & 5.19E-01 & 2.10E+00 & 1.19E+03 & 1.00E+02 & 9.97E+01 & 7.00E+01 & 2.46E-01 \\
      & $10^{11}$ & 5.27E-70 & 2.77E-22 & 2.84E-11 & 1.66E-01 & 1.00E+02 & 1.00E+02 & 9.72E+01 & 8.12E-01 \\
\hline
      & $10^{3}$ & 2.98E-09 & 8.54E-03 & 4.68E-01 & 3.52E+02 & 7.27E+01 & 9.56E+01 & 6.19E+01 & 2.97E-01 \\
$^{140}$Ba & $10^{7}$ & 1.35E-10 & 7.47E-03 & 4.25E-01 & 3.49E+02 & 9.99E+01 & 9.90E+01 & 6.64E+01 & 2.98E-01 \\
      & $10^{11}$ & 7.14E-80 & 2.34E-24 & 4.61E-12 & 4.88E-02 & 1.00E+02 & 1.00E+02 & 9.61E+01 & 9.55E-01 \\
\hline
      & $10^{3}$ & 3.99E-06 & 1.06E-03 & 9.13E-02 & 6.00E+02 & 9.95E+01 & 2.75E+01 & 8.25E-01 & 2.63E-04 \\
$^{142}$Pr & $10^{7}$ & 1.49E-07 & 3.32E-04 & 7.31E-02 & 5.94E+02 & 1.00E+02 & 5.32E+01 & 9.59E-01 & 2.65E-04 \\
      & $10^{11}$ & 1.00E-83 & 2.47E-27 & 9.30E-14 & 8.20E-02 & 1.00E+02 & 7.46E+01 & 3.07E+00 & 6.34E-04 \\
\hline
      & $10^{3}$ & 9.87E-09 & 2.46E-04 & 6.25E-02 & 5.87E+02 & 3.90E+01 & 4.20E+00 & 2.01E-01 & 7.18E-05 \\
$^{198}$Au & $10^{7}$ & 8.05E-11 & 5.49E-05 & 4.99E-02 & 5.83E+02 & 9.91E+01 & 1.44E+01 & 2.39E-01 & 7.19E-05 \\
      & $10^{11}$ & 3.81E-86 & 4.16E-28 & 6.22E-14 & 8.04E-02 & 9.99E+01 & 5.45E+01 & 1.14E+00 & 1.87E-04 \\
\hline
      & $10^{3}$ & 1.04E-05 & 3.53E-03 & 2.75E-01 & 1.21E+03 & 9.97E+01 & 6.01E+01 & 9.13E+00 & 6.56E-03 \\
$^{204}$Au & $10^{7}$ & 1.26E-06 & 1.83E-03 & 2.23E-01 & 1.20E+03 & 1.00E+02 & 8.44E+01 & 1.06E+01 & 6.61E-03 \\
      & $10^{11}$ & 4.79E-77 & 4.37E-25 & 4.64E-13 & 1.66E-01 & 1.00E+02 & 9.97E+01 & 4.64E+01 & 1.79E-02 \\
\hline
      & $10^{3}$ & 2.45E-12 & 2.30E-05 & 1.78E-02 & 3.72E+02 & 1.42E+01 & 2.01E+01 & 5.57E-01 & 1.07E-04 \\
$^{204}$Tl & $10^{7}$ & 1.22E-13 & 7.26E-06 & 1.43E-02 & 3.69E+02 & 9.98E+01 & 4.99E+01 & 6.62E-01 & 1.08E-04 \\
      & $10^{11}$ & 1.22E-87 & 1.29E-28 & 1.79E-14 & 5.07E-02 & 1.00E+02 & 8.87E+01 & 3.18E+00 & 2.82E-04 \\
\hline
\end{tabular}
\end{table}
\onecolumn
\begin{table}
\scriptsize \caption{Same as Table~2 but for electron capture (EC)
direction. Here total rates include $\beta^{+}$-decay and electron
capture rates.}\label{Table 4}

\begin{tabular}{c|c|cccc|cccc}

Nucleus & $\rho$$\it Y_{e}$ & \multicolumn{4}{c|}{$\lambda_{total} (s^{-1})$} & \multicolumn{3}{c}{Percentage contribution of EC} \\
\cline{3-10} & &T$_{9}$=1.5 & T$_{9}$=5 & T$_{9}$=10 & T$_{9}$=30 &T$_{9}$=1.5 & T$_{9}$=5 & T$_{9}$=10 & T$_{9}$=30\\
\hline
     & $10^{3}$ & 1.08E-05 & 5.37E-03 & 3.01E-01 & 4.34E+02 & 6.65E+01 & 9.58E+01 & 9.90E+01 & 1.00E+02 \\
$^{72}$As & $10^{7}$ & 6.30E-03 & 2.35E-02 & 3.75E-01 & 4.37E+02 & 9.99E+01 & 9.90E+01 & 9.92E+01 & 1.00E+02 \\
      & $10^{11}$ & 7.00E+04 & 8.05E+04 & 1.09E+05 & 4.02E+05 & 1.00E+02 & 1.00E+02 & 1.00E+02 & 1.00E+02 \\
\hline
      & $10^{3}$ & 2.23E-08 & 1.15E-03 & 1.69E-01 & 7.43E+02 & 9.97E+01 & 9.96E+01 & 9.96E+01 & 1.00E+02 \\
$^{84}$Rb & $10^{7}$ & 4.06E-05 & 5.34E-03 & 2.10E-01 & 7.50E+02 & 1.00E+02 & 9.99E+01 & 9.97E+01 & 1.00E+02 \\
      & $10^{11}$ & 3.33E+04 & 6.64E+04 & 1.17E+05 & 7.89E+05 & 1.00E+02 & 1.00E+02 & 1.00E+02 & 1.00E+02 \\
\hline
      & $10^{3}$ & 7.36E-08 & 1.44E-03 & 2.63E-01 & 1.11E+03 & 1.00E+02 & 9.96E+01 & 9.98E+01 & 1.00E+02 \\
$^{102}$Rh & $10^{7}$ & 2.69E-04 & 7.00E-03 & 3.29E-01 & 1.12E+03 & 1.00E+02 & 9.99E+01 & 9.99E+01 & 1.00E+02 \\
      & $10^{11}$ & 8.20E+04 & 1.42E+05 & 2.37E+05 & 1.05E+06 & 1.00E+02 & 1.00E+02 & 1.00E+02 & 1.00E+02 \\
\hline
      & $10^{3}$ & 4.35E-09 & 7.21E-04 & 2.40E-01 & 1.29E+03 & 1.00E+02 & 9.98E+01 & 1.00E+02 & 1.00E+02 \\
$^{122}$Sb & $10^{7}$ & 2.00E-05 & 3.37E-03 & 3.00E-01 & 1.30E+03 & 1.00E+02 & 1.00E+02 & 1.00E+02 & 1.00E+02 \\
      & $10^{11}$ & 1.26E+05 & 1.80E+05 & 3.91E+05 & 1.52E+06 & 1.00E+02 & 1.00E+02 & 1.00E+02 & 1.00E+02 \\
\hline
      & $10^{3}$ & 3.23E-04 & 1.64E-02 & 9.30E-01 & 6.93E+02 & 8.79E+00 & 9.05E+01 & 9.89E+01 & 1.00E+02 \\
$^{120}$I & $10^{7}$ & 1.88E-02 & 6.81E-02 & 1.16E+00 & 7.00E+02 & 9.84E+01 & 9.77E+01 & 9.91E+01 & 1.00E+02 \\
      & $10^{11}$ & 3.99E+04 & 6.00E+04 & 1.06E+05 & 3.77E+05 & 1.00E+02 & 1.00E+02 & 1.00E+02 & 1.00E+02 \\
\hline
      & $10^{3}$ & 1.84E-06 & 5.35E-03 & 7.29E-01 & 1.38E+03 & 9.94E+01 & 9.92E+01 & 9.99E+01 & 1.00E+02 \\
$^{124}$I & $10^{7}$ & 1.78E-03 & 2.45E-02 & 9.11E-01 & 1.39E+03 & 1.00E+02 & 9.98E+01 & 9.99E+01 & 1.00E+02 \\
      & $10^{11}$ & 9.77E+04 & 1.58E+05 & 2.85E+05 & 1.01E+06 & 1.00E+02 & 1.00E+02 & 1.00E+02 & 1.00E+02 \\
\hline
      & $10^{3}$ & 1.32E-07 & 1.84E-03 & 3.98E-01 & 1.11E+03 & 1.00E+02 & 9.97E+01 & 1.00E+02 & 1.00E+02 \\
$^{126}$I & $10^{7}$ & 2.64E-04 & 8.68E-03 & 4.98E-01 & 1.12E+03 & 1.00E+02 & 9.99E+01 & 1.00E+02 & 1.00E+02 \\
      & $10^{11}$ & 1.04E+05 & 1.46E+05 & 2.66E+05 & 9.40E+05 & 1.00E+02 & 1.00E+02 & 1.00E+02 & 1.00E+02 \\
\hline
      & $10^{3}$ & 7.13E-05 & 9.51E-03 & 6.76E-01 & 6.56E+02 & 2.79E+01 & 9.57E+01 & 9.95E+01 & 1.00E+02 \\
$^{132}$La & $10^{7}$ & 1.39E-02 & 4.13E-02 & 8.43E-01 & 6.62E+02 & 9.96E+01 & 9.90E+01 & 9.96E+01 & 1.00E+02 \\
      & $10^{11}$ & 3.01E+04 & 4.15E+04 & 8.81E+04 & 3.69E+05 & 1.00E+02 & 1.00E+02 & 1.00E+02 & 1.00E+02 \\
\hline
      & $10^{3}$ & 1.58E-07 & 9.60E-03 & 1.22E+00 & 8.20E+02 & 8.64E+01 & 9.84E+01 & 9.98E+01 & 1.00E+02 \\
$^{198}$Tl & $10^{7}$ & 1.18E-04 & 4.06E-02 & 1.51E+00 & 8.26E+02 & 1.00E+02 & 9.96E+01 & 9.98E+01 & 1.00E+02 \\
      & $10^{11}$ & 4.34E+04 & 5.79E+04 & 1.42E+05 & 3.79E+05 & 1.00E+02 & 1.00E+02 & 1.00E+02 & 1.00E+02 \\
\hline
\end{tabular}
\end{table}
\onecolumn
\begin{table}
\scriptsize \caption {Same as Table~2 but for U1F
rates.}\label{Table 5}

\begin{tabular}{c|c|cccc|cccc}

Nucleus & $\rho$$\it Y_{e}$ & \multicolumn{4}{c|}{$\lambda_{total} (s^{-1})$} & \multicolumn{3}{c}{Percentage contribution of $\beta^{-}$-decay} \\
\cline{3-10} & &T$_{9}$=1.5 & T$_{9}$=5 & T$_{9}$=10 & T$_{9}$=30 &T$_{9}$=1.5 & T$_{9}$=5 & T$_{9}$=10 & T$_{9}$=30\\
\hline
      & $10^{3}$ & 7.08E-02 & 3.59E-02 & 5.46E-01 & 3.48E+04 & 1.00E+02 & 9.66E+01 & 2.93E+00 & 9.44E-06 \\
$^{82}$Br & $10^{7}$ & 6.46E-02 & 3.28E-02 & 4.37E-01 & 3.46E+04 & 1.00E+02 & 9.93E+01 & 3.57E+00 & 9.48E-06 \\
      & $10^{11}$ & 1.46E-72 & 8.55E-24 & 7.64E-13 & 4.75E+00 & 1.00E+02 & 1.00E+02 & 3.24E+01 & 3.15E-05 \\
\hline
      & $10^{3}$ & 4.36E-01 & 1.72E-01 & 1.63E+00 & 5.35E+04 & 1.00E+02 & 9.83E+01 & 4.53E+00 & 2.86E-05 \\
$^{84}$Br & $10^{7}$ & 4.24E-01 & 1.67E-01 & 1.31E+00 & 5.31E+04 & 1.00E+02 & 9.97E+01 & 5.55E+00 & 2.87E-05 \\
      & $10^{11}$ & 5.81E-67 & 7.11E-22 & 5.34E-12 & 7.31E+00 & 1.00E+02 & 1.00E+02 & 7.16E+01 & 1.26E-04 \\
\hline
      & $10^{3}$ & 5.01E+00 & 1.75E+00 & 7.52E+00 & 8.93E+04 & 1.00E+02 & 9.74E+01 & 1.07E+01 & 2.18E-04 \\
$^{86}$Br & $10^{7}$ & 4.99E+00 & 1.70E+00 & 6.16E+00 & 8.87E+04 & 1.00E+02 & 9.95E+01 & 1.30E+01 & 2.19E-04 \\
      & $10^{11}$ & 1.82E-56 & 2.35E-18 & 5.46E-10 & 1.22E+01 & 1.00E+02 & 1.00E+02 & 9.88E+01 & 1.77E-03 \\
\hline
      & $10^{3}$ & 1.64E-02 & 3.83E-02 & 3.11E+00 & 2.78E+04 & 1.00E+02 & 4.20E+01 & 4.21E-01 & 1.14E-05 \\
$^{88}$Kr & $10^{7}$ & 1.48E-02 & 1.94E-02 & 2.48E+00 & 2.75E+04 & 1.00E+02 & 7.72E+01 & 5.14E-01 & 1.15E-05 \\
      & $10^{11}$ & 9.95E-74 & 2.94E-24 & 3.23E-12 & 3.81E+00 & 1.00E+02 & 9.94E+01 & 5.50E+00 & 3.62E-05 \\
\hline
      & $10^{3}$ & 4.53E-04 & 7.59E-04 & 1.38E-01 & 2.07E+04 & 9.97E+01 & 1.76E+01 & 3.87E-02 & 6.05E-08 \\
$^{84}$Rb & $10^{7}$ & 7.94E-05 & 2.18E-04 & 1.10E-01 & 2.05E+04 & 1.00E+02 & 4.23E+01 & 4.55E-02 & 6.08E-08 \\
      & $10^{11}$ & 4.86E-81 & 1.57E-27 & 1.35E-13 & 2.82E+00 & 1.00E+02 & 6.77E+01 & 1.51E-01 & 1.46E-07 \\
\hline
      & $10^{3}$ & 7.13E-03 & 2.87E-03 & 2.39E-01 & 2.45E+04 & 1.00E+02 & 7.77E+01 & 3.97E-01 & 8.41E-07 \\
$^{86}$Rb & $10^{7}$ & 4.75E-03 & 2.00E-03 & 1.90E-01 & 2.43E+04 & 1.00E+02 & 9.36E+01 & 4.77E-01 & 8.47E-07 \\
      & $10^{11}$ & 1.92E-77 & 6.29E-26 & 2.38E-13 & 3.35E+00 & 1.00E+02 & 9.92E+01 & 2.51E+00 & 2.29E-06 \\
\hline
      & $10^{3}$ & 1.50E+00 & 6.60E-01 & 2.77E+00 & 5.85E+04 & 1.00E+02 & 9.92E+01 & 1.17E+01 & 1.34E-04 \\
$^{88}$Rb & $10^{7}$ & 1.48E+00 & 6.45E-01 & 2.27E+00 & 5.81E+04 & 1.00E+02 & 9.98E+01 & 1.42E+01 & 1.35E-04 \\
      & $10^{11}$ & 2.28E-64 & 9.48E-21 & 3.14E-11 & 8.00E+00 & 1.00E+02 & 1.00E+02 & 9.24E+01 & 6.71E-04 \\
\hline
      & $10^{3}$ & 6.06E-06 & 4.37E-03 & 2.47E+00 & 4.02E+04 & 9.85E+01 & 1.15E-01 & 1.51E-04 & 2.15E-09 \\
$^{90}$Sr & $10^{7}$ & 2.62E-07 & 8.68E-04 & 1.96E+00 & 3.99E+04 & 1.00E+02 & 3.54E-01 & 1.76E-04 & 2.16E-09 \\
      & $10^{11}$ & 1.14E-40 & 3.11E-14 & 2.41E-12 & 5.51E+00 & 1.00E+02 & 7.58E-01 & 5.05E-04 & 4.95E-09 \\
\hline
      & $10^{3}$ & 2.01E-04 & 1.67E-03 & 6.31E-01 & 7.78E+03 & 1.00E+02 & 1.15E+01 & 3.33E-02 & 3.70E-06 \\
$^{92}$Sr & $10^{7}$ & 1.45E-04 & 4.57E-04 & 5.06E-01 & 7.73E+03 & 1.00E+02 & 3.60E+01 & 4.01E-02 & 3.72E-06 \\
      & $10^{11}$ & 1.67E-78 & 8.11E-27 & 7.11E-13 & 1.07E+00 & 1.00E+02 & 8.55E+01 & 2.26E-01 & 9.59E-06 \\
\hline
      & $10^{3}$ & 1.55E-02 & 6.65E-03 & 4.27E-01 & 3.81E+04 & 1.00E+02 & 9.15E+01 & 6.68E-01 & 1.89E-06 \\
$^{90}$Y & $10^{7}$ & 1.25E-02 & 5.53E-03 & 3.41E-01 & 3.78E+04 & 1.00E+02 & 9.79E+01 & 8.12E-01 & 1.90E-06 \\
      & $10^{11}$ & 1.17E-75 & 3.77E-25 & 4.39E-13 & 5.20E+00 & 1.00E+02 & 9.99E+01 & 5.67E+00 & 5.53E-06 \\
\hline

\end{tabular}
\end{table}
\onecolumn
\begin{table}
\scriptsize \caption{Same as Table~2 but for U1F rates.}\label{Table
6}

\begin{tabular}{c|c|cccc|cccc}

Nucleus & $\rho$$\it Y_{e}$ & \multicolumn{4}{c|}{$\lambda_{total} (s^{-1})$} & \multicolumn{3}{c}{Percentage contribution of $\beta^{-}$-decay} \\
\cline{3-10} & &T$_{9}$=1.5 & T$_{9}$=5 & T$_{9}$=10 & T$_{9}$=30 &T$_{9}$=1.5 & T$_{9}$=5 & T$_{9}$=10 & T$_{9}$=30\\
\hline
      & $10^{3}$ & 3.72E-01 & 1.80E-01 & 1.20E+00 & 5.19E+04 & 1.00E+02 & 9.86E+01 & 7.45E+00 & 3.48E-05 \\
$^{92}$Y & $10^{7}$ & 3.51E-01 & 1.71E-01 & 9.75E-01 & 5.15E+04 & 1.00E+02 & 9.97E+01 & 9.02E+00 & 3.49E-05 \\
      & $10^{11}$ & 3.58E-70 & 1.15E-22 & 3.17E-12 & 7.10E+00 & 1.00E+02 & 1.00E+02 & 6.57E+01 & 1.27E-04 \\
\hline
      & $10^{3}$ & 1.53E-01 & 7.81E-02 & 1.25E+00 & 6.25E+04 & 1.00E+02 & 9.85E+01 & 2.83E+00 & 1.22E-05 \\
$^{94}$Y & $10^{7}$ & 1.50E-01 & 7.56E-02 & 1.01E+00 & 6.19E+04 & 1.00E+02 & 9.97E+01 & 3.50E+00 & 1.23E-05 \\
      & $10^{11}$ & 1.34E-66 & 5.25E-22 & 3.47E-12 & 8.53E+00 & 1.00E+02 & 1.00E+02 & 6.57E+01 & 5.66E-05 \\
\hline
      & $10^{3}$ & 2.89E-05 & 2.94E-05 & 1.10E-01 & 2.79E+04 & 9.99E+01 & 3.71E+01 & 4.10E-03 & 3.66E-09 \\
$^{102}$Rh & $10^{7}$ & 9.59E-06 & 1.18E-05 & 8.73E-02 & 2.77E+04 & 1.00E+02 & 6.86E+01 & 4.86E-03 & 3.68E-09 \\
      & $10^{11}$ & 1.43E-81 & 1.36E-28 & 1.07E-13 & 3.80E+00 & 1.00E+02 & 8.90E+01 & 1.84E-02 & 9.14E-09 \\
\hline
      & $10^{3}$ & 4.04E-04 & 1.62E-04 & 4.97E-01 & 6.90E+04 & 1.00E+02 & 7.38E+01 & 9.84E-03 & 1.69E-08 \\
$^{122}$Sb & $10^{7}$ & 2.92E-04 & 1.11E-04 & 3.95E-01 & 6.85E+04 & 1.00E+02 & 9.24E+01 & 1.19E-02 & 1.70E-08 \\
      & $10^{11}$ & 4.05E-78 & 4.53E-27 & 4.85E-13 & 9.42E+00 & 1.00E+02 & 9.93E+01 & 7.16E-02 & 4.72E-08 \\
\hline
      & $10^{3}$ & 7.78E-01 & 2.45E-01 & 1.18E+01 & 1.15E+05 & 1.00E+02 & 8.57E+01 & 7.74E-01 & 2.14E-05 \\
$^{136}$I & $10^{7}$ & 7.71E-01 & 2.15E-01 & 9.45E+00 & 1.14E+05 & 1.00E+02 & 9.68E+01 & 9.63E-01 & 2.15E-05 \\
      & $10^{11}$ & 1.09E-59 & 6.30E-20 & 4.22E-11 & 1.57E+01 & 1.00E+02 & 1.00E+02 & 7.28E+01 & 1.47E-04 \\
\hline
      & $10^{3}$ & 7.47E-04 & 4.53E-03 & 4.40E+00 & 8.28E+04 & 9.99E+01 & 1.45E+01 & 1.02E-02 & 1.11E-07 \\
$^{140}$Ba & $10^{7}$ & 1.78E-04 & 1.23E-03 & 3.51E+00 & 8.20E+04 & 1.00E+02 & 3.77E+01 & 1.20E-02 & 1.12E-07 \\
      & $10^{11}$ & 1.58E-80 & 9.06E-27 & 4.30E-12 & 1.13E+01 & 1.00E+02 & 6.59E+01 & 4.20E-02 & 2.70E-07 \\
\hline
      & $10^{3}$ & 2.79E-02 & 9.87E-03 & 1.27E+00 & 9.16E+04 & 1.00E+02 & 9.52E+01 & 3.09E-01 & 1.09E-06 \\
$^{142}$Pr & $10^{7}$ & 2.14E-02 & 8.32E-03 & 1.01E+00 & 9.08E+04 & 1.00E+02 & 9.89E+01 & 3.74E-01 & 1.10E-06 \\
      & $10^{11}$ & 9.20E-76 & 4.68E-25 & 1.26E-12 & 1.25E+01 & 1.00E+02 & 9.99E+01 & 2.46E+00 & 3.12E-06 \\
\hline
      & $10^{3}$ & 5.57E-06 & 2.33E-05 & 6.61E-01 & 8.36E+04 & 1.00E+02 & 6.50E+00 & 9.42E-05 & 2.14E-10 \\
$^{198}$Au & $10^{7}$ & 2.40E-06 & 5.47E-06 & 5.26E-01 & 8.30E+04 & 1.00E+02 & 2.11E+01 & 1.12E-04 & 2.14E-10 \\
      & $10^{11}$ & 9.75E-82 & 3.89E-29 & 6.44E-13 & 1.14E+01 & 1.00E+02 & 5.54E+01 & 4.63E-04 & 5.43E-10 \\
\hline
      & $10^{3}$ & 3.72E+00 & 1.62E+00 & 3.19E+00 & 9.29E+04 & 1.00E+02 & 9.98E+01 & 1.92E+01 & 1.32E-04 \\
$^{204}$Au & $10^{7}$ & 3.52E+00 & 1.56E+00 & 2.65E+00 & 9.23E+04 & 1.00E+02 & 1.00E+02 & 2.27E+01 & 1.33E-04 \\
      & $10^{11}$ & 5.42E-68 & 2.02E-21 & 2.10E-11 & 1.27E+01 & 1.00E+02 & 1.00E+02 & 8.81E+01 & 5.12E-04 \\
\hline
      & $10^{3}$ & 4.41E-05 & 2.69E-05 & 2.71E-01 & 5.31E+04 & 1.00E+02 & 4.97E+01 & 1.75E-03 & 2.11E-09 \\
$^{204}$Tl & $10^{7}$ & 4.10E-06 & 1.13E-05 & 2.16E-01 & 5.27E+04 & 1.00E+02 & 7.61E+01 & 2.05E-03 & 2.12E-09 \\
      & $10^{11}$ & 1.88E-82 & 9.52E-29 & 2.64E-13 & 7.24E+00 & 1.00E+02 & 8.85E+01 & 6.27E-03 & 4.98E-09 \\
\hline
\end{tabular}
\end{table}

\onecolumn
\begin{table}
\scriptsize \caption{Same as Table~4 but for U1F rates.}\label{Table
7}

\begin{tabular}{c|c|cccc|cccc}

Nucleus & $\rho$$\it Y_{e}$ & \multicolumn{4}{c|}{$\lambda_{total} (s^{-1})$} & \multicolumn{3}{c}{Percentage contribution of EC} \\
\cline{3-10} & &T$_{9}$=1.5 & T$_{9}$=5 & T$_{9}$=10 & T$_{9}$=30 &T$_{9}$=1.5 & T$_{9}$=5 & T$_{9}$=10 & T$_{9}$=30\\
\hline
      & $10^{3}$ & 6.34E-03 & 5.78E-03 & 1.24E+00 & 4.06E+04 & 9.30E-02 & 4.75E+01 & 9.99E+01 & 1.00E+02 \\
$^{72}$As & $10^{7}$ & 1.31E-02 & 1.63E-02 & 1.55E+00 & 4.09E+04 & 5.16E+01 & 8.11E+01 & 9.99E+01 & 1.00E+02 \\
      & $10^{11}$ & 5.64E+06 & 6.58E+06 & 9.55E+06 & 7.62E+07 & 1.00E+02 & 1.00E+02 & 1.00E+02 & 1.00E+02 \\
\hline
      & $10^{3}$ & 9.49E-05 & 4.34E-04 & 1.26E-01 & 1.99E+04 & 1.84E+00 & 9.32E+01 & 1.00E+02 & 1.00E+02 \\
$^{84}$Rb & $10^{7}$ & 2.18E-03 & 1.93E-03 & 1.58E-01 & 2.01E+04 & 9.57E+01 & 9.84E+01 & 1.00E+02 & 1.00E+02 \\
      & $10^{11}$ & 2.28E+06 & 4.00E+06 & 6.19E+06 & 5.18E+07 & 1.00E+02 & 1.00E+02 & 1.00E+02 & 1.00E+02 \\
\hline
      & $10^{3}$ & 9.49E-05 & 4.34E-04 & 3.27E+00 & 1.26E+05 & 1.84E+00 & 9.32E+01 & 1.00E+02 & 1.00E+02 \\
$^{102}$Rh & $10^{7}$ & 2.18E-03 & 1.93E-03 & 4.10E+00 & 1.26E+05 & 9.57E+01 & 9.84E+01 & 1.00E+02 & 1.00E+02 \\
      & $10^{11}$ & 2.28E+06 & 4.00E+06 & 2.84E+07 & 2.26E+08 & 1.00E+02 & 1.00E+02 & 1.00E+02 & 1.00E+02 \\
\hline
      & $10^{3}$ & 1.76E-05 & 3.31E-03 & 1.60E+00 & 1.29E+05 & 7.61E+01 & 1.00E+02 & 1.00E+02 & 1.00E+02 \\
$^{122}$Sb & $10^{7}$ & 1.63E-02 & 1.57E-02 & 2.01E+00 & 1.30E+05 & 1.00E+02 & 1.00E+02 & 1.00E+02 & 1.00E+02 \\
      & $10^{11}$ & 1.16E+07 & 1.67E+07 & 4.05E+07 & 2.86E+08 & 1.00E+02 & 1.00E+02 & 1.00E+02 & 1.00E+02 \\
\hline
      & $10^{3}$ & 3.14E-02 & 2.95E-01 & 9.40E+01 & 5.56E+05 & 1.11E-01 & 9.36E+01 & 1.00E+02 & 1.00E+02 \\
$^{120}$I & $10^{7}$ & 8.10E-02 & 1.39E+00 & 1.18E+02 & 5.61E+05 & 6.13E+01 & 9.86E+01 & 1.00E+02 & 1.00E+02 \\
      & $10^{11}$ & 3.28E+07 & 7.31E+07 & 1.50E+08 & 7.52E+08 & 1.00E+02 & 1.00E+02 & 1.00E+02 & 1.00E+02 \\
\hline
      & $10^{3}$ & 1.22E-03 & 1.49E-02 & 1.36E+01 & 2.25E+05 & 1.67E+00 & 9.70E+01 & 1.00E+02 & 1.00E+02 \\
$^{124}$I & $10^{7}$ & 2.22E-02 & 7.17E-02 & 1.71E+01 & 2.28E+05 & 9.46E+01 & 9.94E+01 & 1.00E+02 & 1.00E+02 \\
      & $10^{11}$ & 1.49E+07 & 2.66E+07 & 5.60E+07 & 3.70E+08 & 1.00E+02 & 1.00E+02 & 1.00E+02 & 1.00E+02 \\
\hline
      & $10^{3}$ & 6.87E-05 & 4.18E-03 & 5.78E+00 & 1.63E+05 & 1.82E+01 & 9.96E+01 & 1.00E+02 & 1.00E+02 \\
$^{126}$I & $10^{7}$ & 1.40E-02 & 2.01E-02 & 7.26E+00 & 1.65E+05 & 9.96E+01 & 9.99E+01 & 1.00E+02 & 1.00E+02 \\
      & $10^{11}$ & 1.51E+07 & 2.02E+07 & 4.37E+07 & 3.02E+08 & 1.00E+02 & 1.00E+02 & 1.00E+02 & 1.00E+02 \\
\hline
      & $10^{3}$ & 2.97E-03 & 1.16E-01 & 3.72E+01 & 2.85E+05 & 5.45E-01 & 9.91E+01 & 1.00E+02 & 1.00E+02 \\
$^{132}$La & $10^{7}$ & 3.17E-02 & 5.67E-01 & 4.66E+01 & 2.88E+05 & 9.07E+01 & 9.98E+01 & 1.00E+02 & 1.00E+02 \\
      & $10^{11}$ & 1.85E+07 & 2.65E+07 & 6.07E+07 & 4.06E+08 & 1.00E+02 & 1.00E+02 & 1.00E+02 & 1.00E+02 \\
\hline
      & $10^{3}$ & 5.62E-04 & 6.64E-03 & 1.21E+01 & 1.78E+05 & 4.91E+00 & 9.74E+01 & 1.00E+02 & 1.00E+02 \\
$^{198}$Tl & $10^{7}$ & 2.31E-02 & 3.14E-02 & 1.52E+01 & 1.79E+05 & 9.77E+01 & 9.94E+01 & 1.00E+02 & 1.00E+02 \\
      & $10^{11}$ & 1.41E+07 & 1.80E+07 & 5.58E+07 & 3.06E+08 & 1.00E+02 & 1.00E+02 & 1.00E+02 & 1.00E+02 \\
\hline
\end{tabular}
\end{table}

\clearpage


\begin{figure}[h]
\includegraphics[scale=0.5]{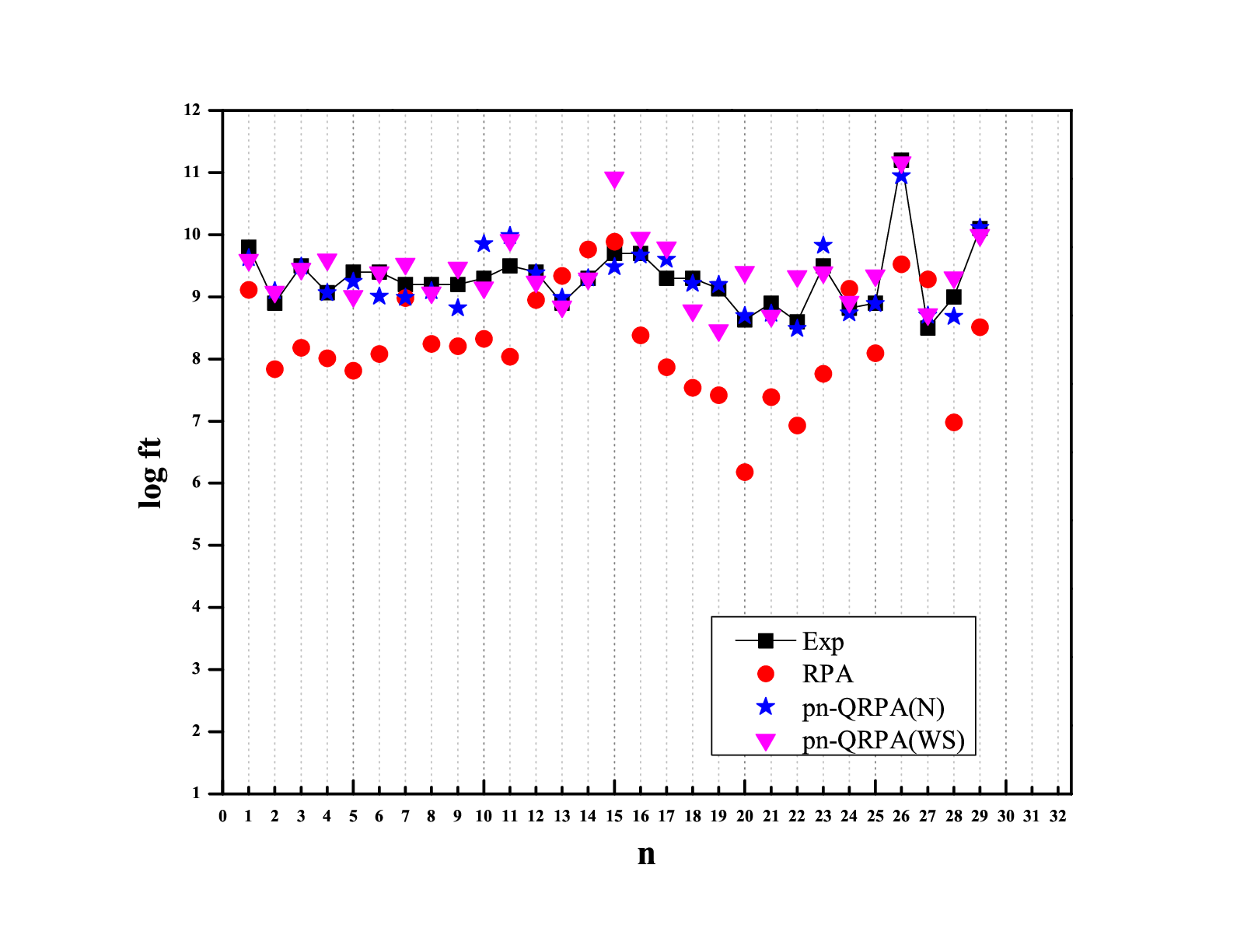}
\normalsize \caption {Experimental and theoretical $log \emph{ft}$
values for $\Delta J = 2$ unique first-forbidden $\beta-decay$
transitions. The transitions are ordered in the same way as in Table
1. The solid line represents the experimental data \cite{Led78}. RPA
results are taken from \cite{Civ86}.}\label{fig1}
\end{figure}

\begin{figure}[h]
\begin{tabular}{cc}
\includegraphics[scale=0.37]{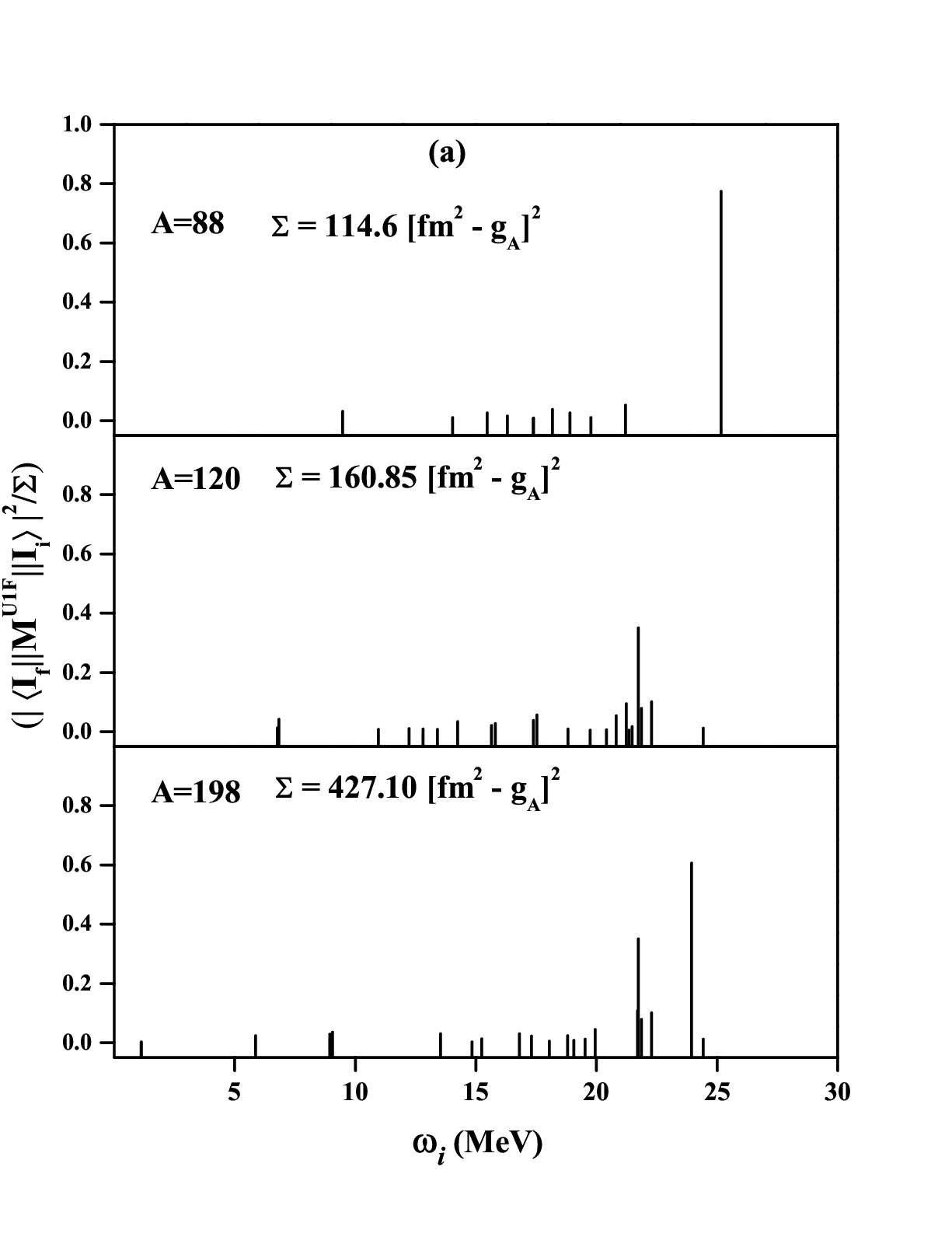}
\includegraphics[scale=0.36]{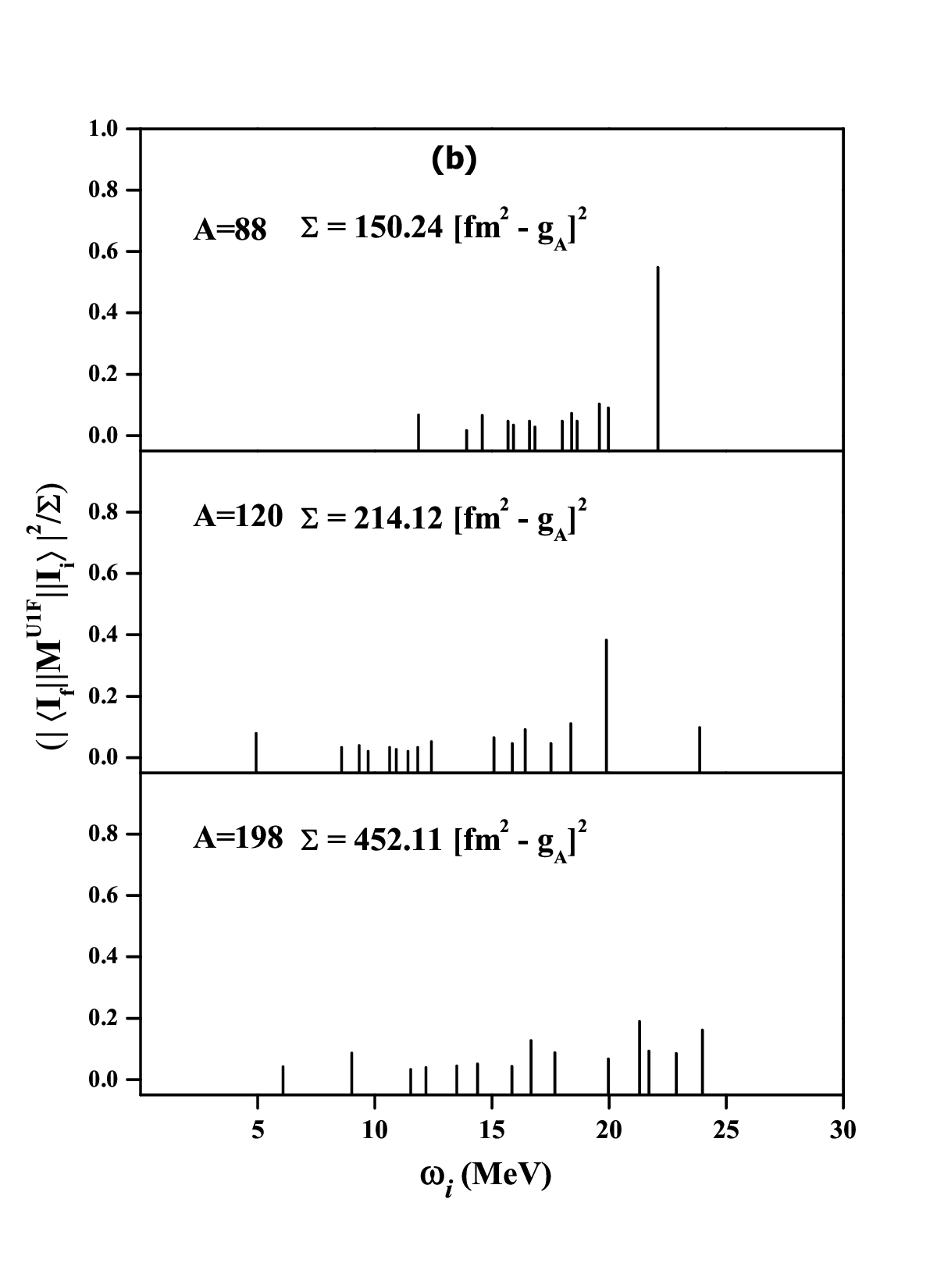}
\end{tabular}
\normalsize \caption{Comparison of strength distributions for a few
$2^{-}\longrightarrow 0^{+}$ transitions as a function of daughter
excitation energy ($\omega_{i}$) and in units of the total strength
$\Sigma$. (a) pn-QRPA(WS) matrix elements, (b) correlated RPA
\cite{Civ86} matrix elements.}\label{fig2}
\end{figure}

\begin{figure}[h]
\includegraphics[scale=0.3]{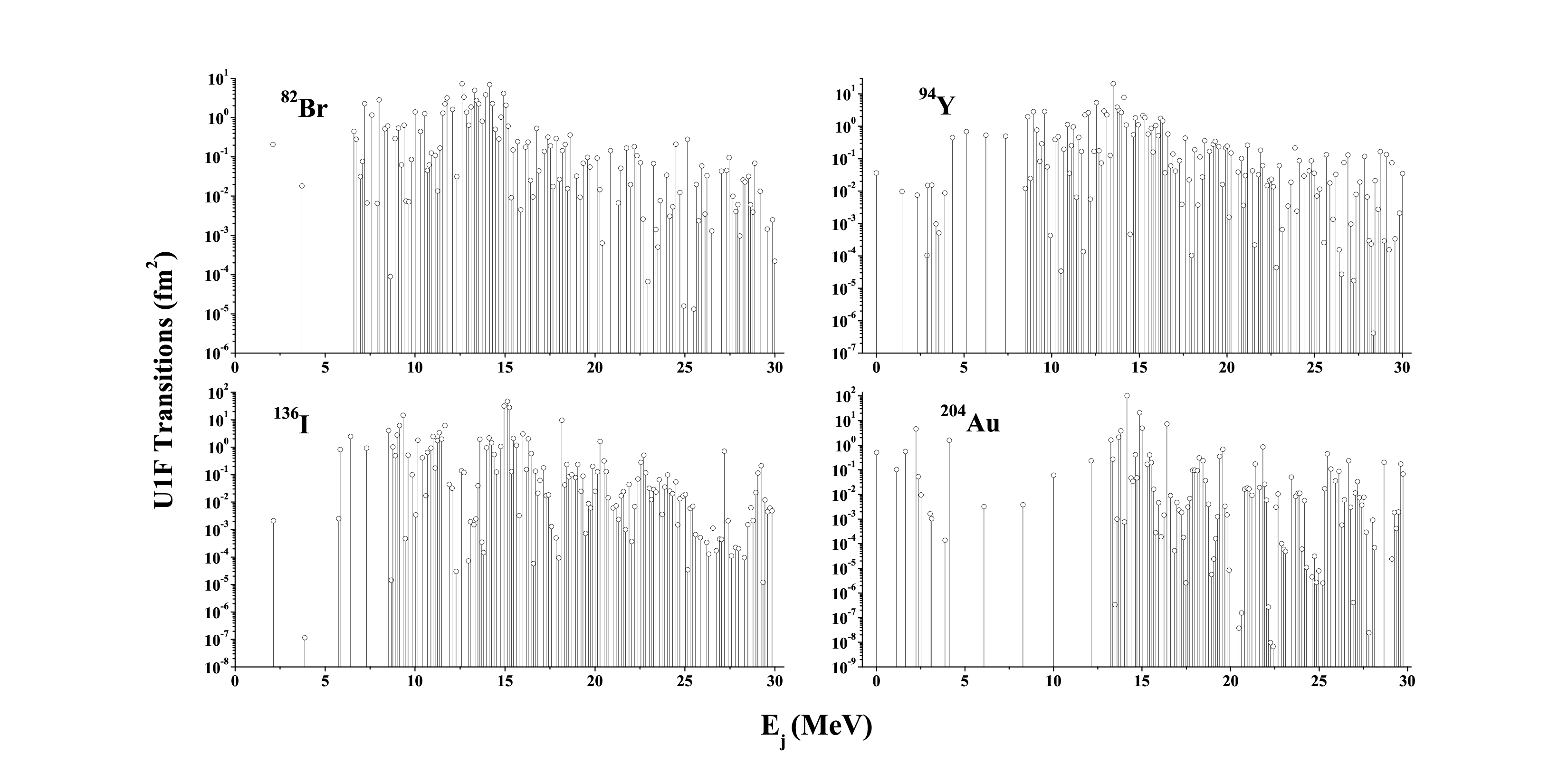}
\normalsize \caption {Calculated U1F transitions for selected nuclei
in $\beta$-decay direction using the pn-QRPA(N) model.}\label{fig3}
\end{figure}

\begin{figure}[h]
\includegraphics[scale=0.3]{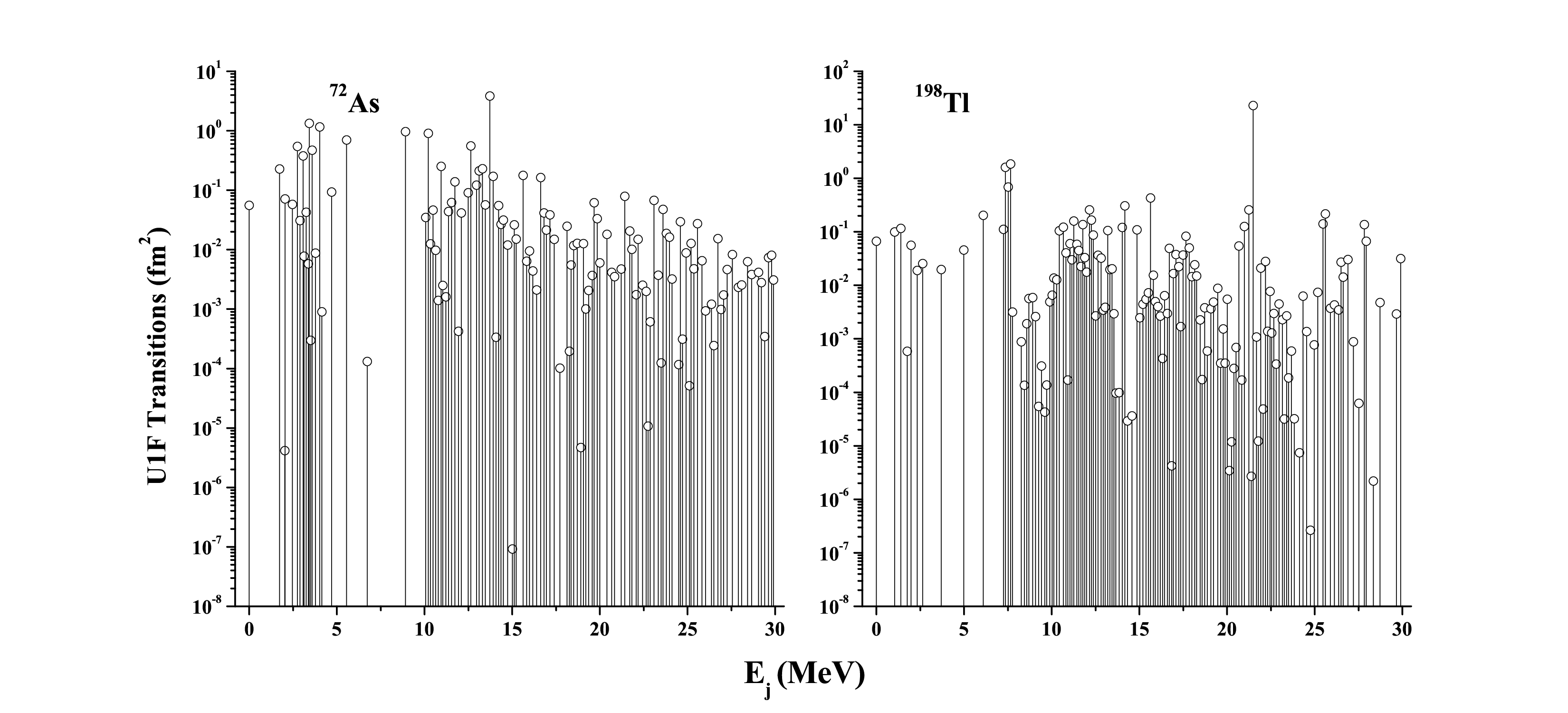}
\normalsize \caption {Calculated U1F transitions for selected nuclei
in electron capture direction using the pn-QRPA(N)
model.}\label{fig4}
\end{figure}

\begin{figure}[h]
\includegraphics[scale=0.45]{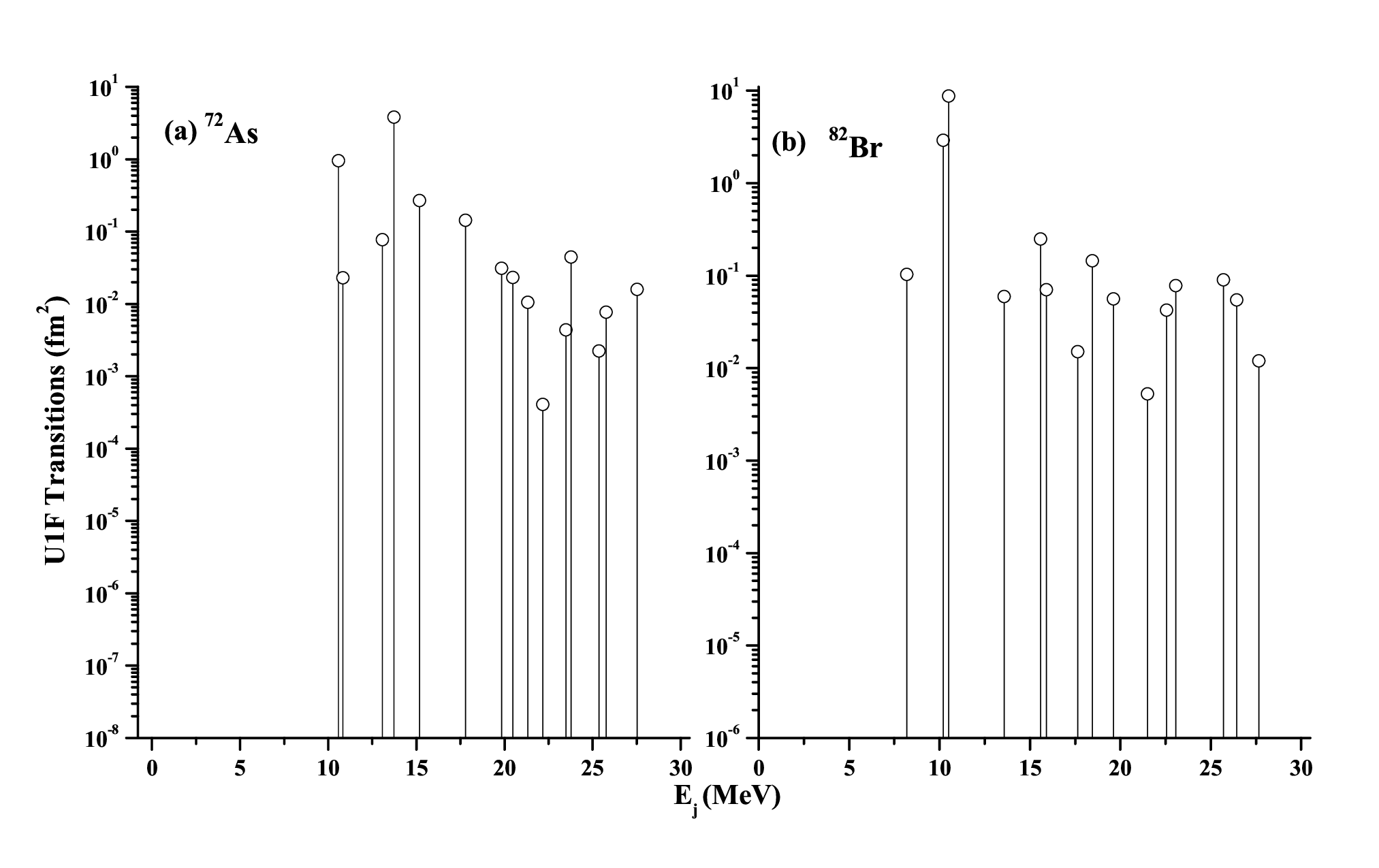}
\normalsize \caption {Calculated U1F transitions for selected nuclei
in (a) electron capture (b) $\beta$-decay direction using the
pn-QRPA(N) model in spherical limit (without deformation of
nuclei).}\label{fig5}
\end{figure}

\begin{figure}[h]
\includegraphics[scale=0.35]{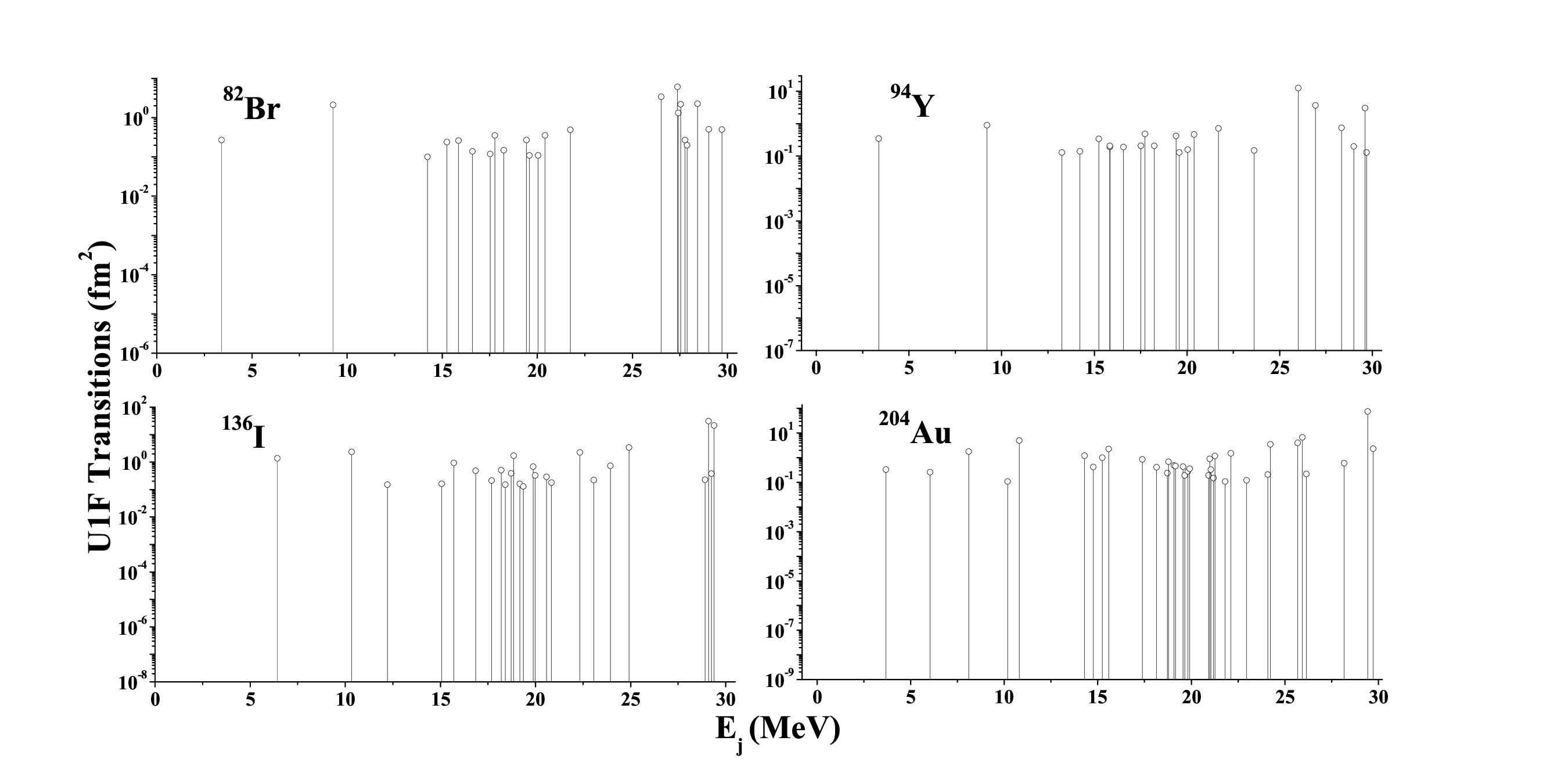}
\normalsize \caption {Calculated U1F transitions for selected nuclei
in $\beta$-decay direction using the pn-QRPA(WS) model.}\label{fig6}
\end{figure}

\begin{figure}[h]
\includegraphics[scale=0.35]{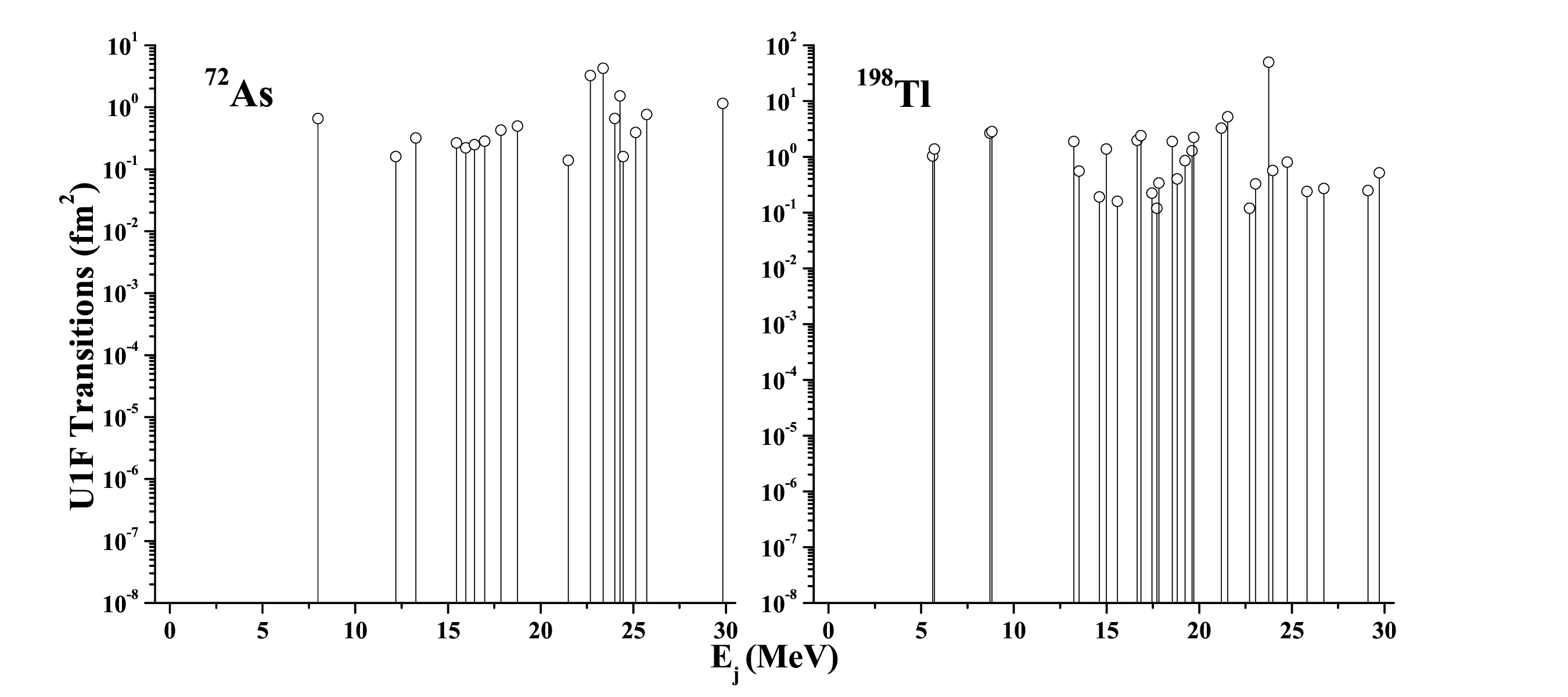}
\normalsize \caption {Calculated U1F transitions for selected nuclei
in electron capture direction using the pn-QRPA(WS)
model.}\label{fig7}
\end{figure}

\begin{figure}[h]
\includegraphics[scale=0.52]{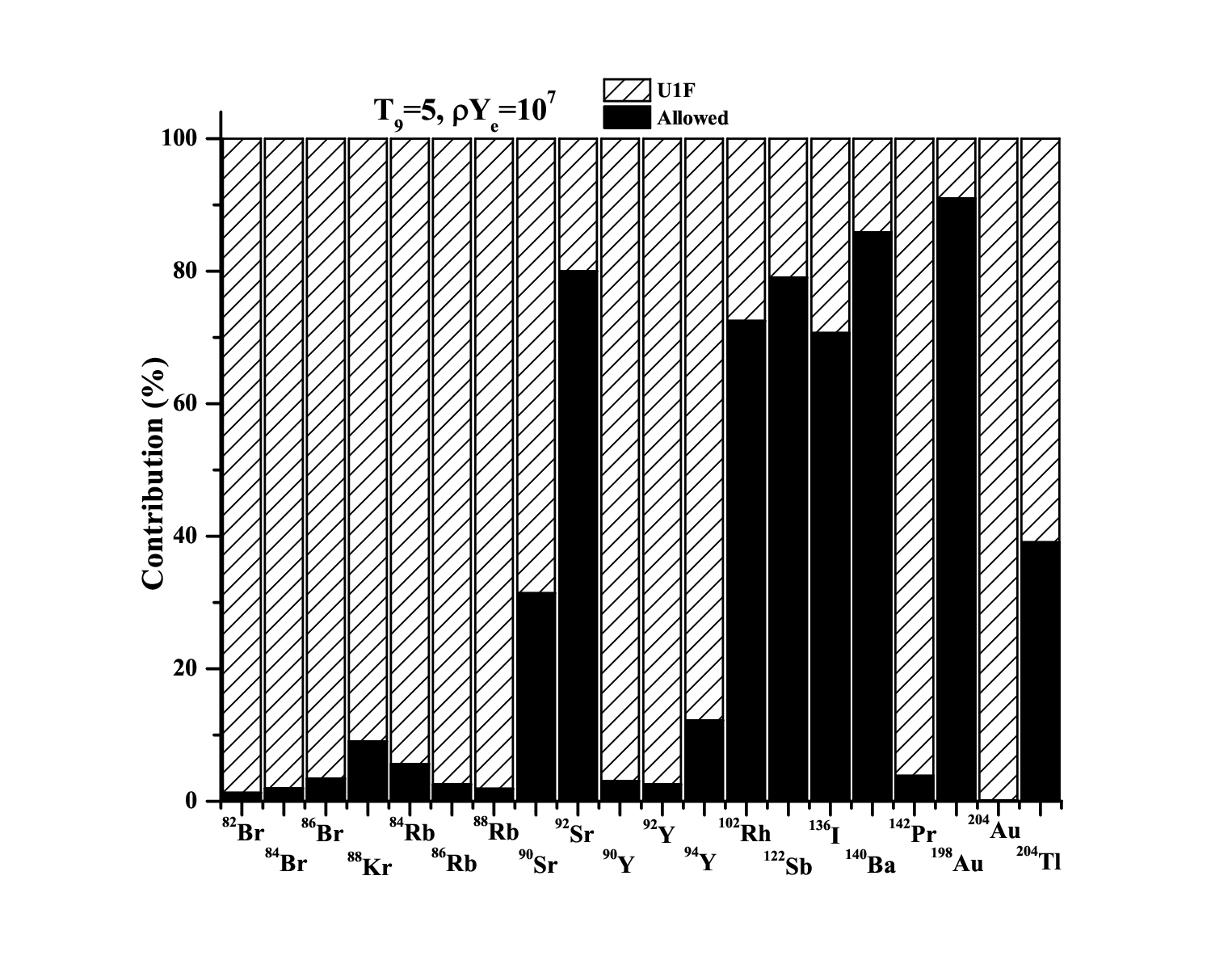}
\normalsize \caption {Contribution of allowed and U1F transitions to
total rates in $\beta$-decay direction. Stellar density
($\rho$Y$_{e}$) is given in units of g/cm$^{3}$, whereas temperature
(T$_{9}$) is given in units of 10$^{9}$ K.}\label{fig8}
\end{figure}

\begin{figure}[h]
\includegraphics[scale=0.55]{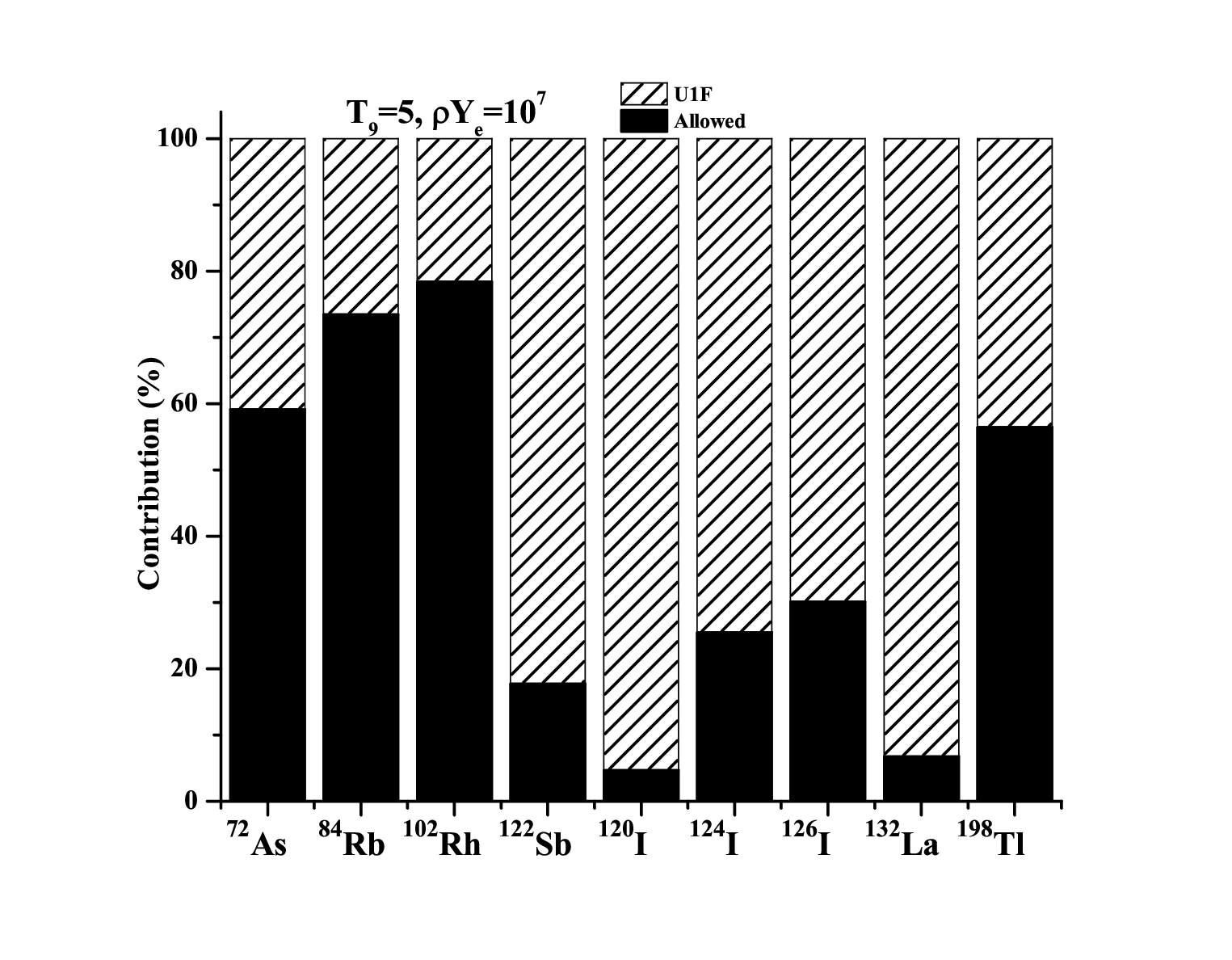}
\normalsize \caption{Contribution of allowed and U1F transitions to
total rates in EC direction. Stellar density ($\rho$Y$_{e}$) is
given in units of g/cm$^{3}$, whereas temperature (T$_{9}$) is given
in units of 10$^{9}$ K.}\label{fig9}
\end{figure}


\begin{thebibliography}{48}
\bibitem{Cow04}Cowan J J and Thielemann F-K, Phys. Today, 57/10  47 (2004).
\bibitem{Bor06}Borzov I N \emph{Nucl. Phys.} A 777 645-675 (2006).
\bibitem{Nabi10}Nabi J.-U \emph{Advances in Space Research} 46 1191-1207 (2010).
\bibitem{Zhi13}Zhi Q \emph{et al.}, \emph{Phys. Rev.} C  87 025803 (2013).
\bibitem{Civ96}Civitarese O and Suhonen J \emph{Nucl. Phys.} A 607 152-162 (1996).
\bibitem{Tre95}Tretyak V I and Zdesenko Y G \emph{Atomic and Nucl. Data Tables} 61 43-90 (1995).
\bibitem{Moe94}Moe M and Vogel P \emph{Ann. Rev. Nucl. Part. Sci.} 44 247-283 (1994).
\bibitem{Tak69}Takahashi K and Yamada M \emph{Prog. Theor. Phys.} 41 1470-1503 (1969).
\bibitem{Hal67}Halbleib J A and Sorensen R A \emph{Nucl. Phys.} A 98 542 (1967).
\bibitem{Ran73}Randrup J \emph{Nucl. Phys.} A 207 209 (1973).
\bibitem{Gab73}Gabrakov S I, Kuliev A A and Salamov D I, \emph{Preprint IGTP IV} 73, 166 (1973).
\bibitem{Gab76}Gabrakov S I, Kuliev A A and Salamov D I, \emph{Sov. J. Nuc. Phys.} 24 (1976).
\bibitem{Iva77}Ivanova S P, Kuliev A A and Salamov D I, \emph{Bulletin of Academy of Sci. of the USSR ser.Phys} pp.131-138 (1977).
\bibitem{Mut89}Muto K \emph{et al.}, \emph{Z. Phys.} A 333 125 (1989).
\bibitem{Mol90}M$\ddot{o}$ller P and Randrup J \emph{Nucl. Phys.} A 514 1 (1990).
\bibitem{Kla84}Klapdor H V, Metzinger J and Oda T \emph{At. Data Nucl. Data Tables} 31 81 (1984).
\bibitem{Sta90}Hirsch M, Staudt A, Muto K and Klapdor-Kleingrothaus H V \emph{At. Data Nucl. Data Tables} 53 165 (1993).
\bibitem{Hir93}Staudt A, Bender E, Muto K and Klapdor-Kleingrothaus H V \emph{At. Data Nucl. Data Tables} 44 79 (1993).
\bibitem{Hom96}Homma H \emph{et al.}, \emph{Phys. Rev.} C 54 2972 (1996).
\bibitem{Ken05}Kenar I, Selam C and Kucukbursa A, \emph{Math. Comp. Applications} 10 2 (2005).
\bibitem{Nab99a}Nabi J.-U and Klapdor-Kleingrothaus H V \emph{Eur. Phys. J. A } 5 337 (1999).
\bibitem{Nab99}Nabi J.-U and Klapdor-Kleingrothaus H V \emph{At. Data Nucl. Data Tables} 71 149 (1999).
\bibitem{Nab04}Nabi J.-U and Klapdor-Kleingrothaus H V \emph{At. Data Nucl. Data Tables} 88 237 (2004).
\bibitem{Nab14}Nabi J.-U and Stoica S \emph{Astrophys. Space Sci.} 349 843-855 (2014).
\bibitem{Suh93}Suhonen J \emph{Nucl. Phys.} A 563, 205 (1993).
\bibitem{Civ86}Civitarese O \emph{et al.}, \emph{Nucl. Phys.} A 453 45-57 (1986).
\bibitem{Nec10}Cakmak N \emph{et al.}, \emph{Pramana J. Phys.} 74 541-553 (2010).
\bibitem{Suz12}Suzuki T \emph{et al.}, \emph{Phys. Rev.} C 85 015802 (2012).
\bibitem{Mar16}Marketin T \emph{et al.}, \emph{Phys. Rev.} C 93 025805 (2016).
\bibitem{Sol76}Soloviev V G \emph{Theory of Complex Nuclei, Pergamon, New York} (1976).
\bibitem{Mut92}Muto K \emph{et al.}, \emph{Z. Phys.} A 341  407 (1992).
\bibitem{Boh69}Bohr A and Mottelson B R \emph{Nuclear Structure, Benjamin, W.A. Inc., New York} Vol. 1 (1969).
\bibitem{Nab16}Nabi J.-U, Cakmak N and Iftikhar Z \emph{Eur. Phys. J. A } 52 1 (2016).
\bibitem{Eng88}Engel J, Vogel P and Zirnbauer M R \emph{Phys. Rev.} C 37 731 (1988).
\bibitem{Mut89a}Muto K, Bender E and Klapdor H V \emph{Z. Phys.} A 334 47 (1989).
\bibitem{Mol81}M\"{o}ller P and Nix J R \emph{At. Data Nucl. Data Tables} 26 165 (1981).
\bibitem{Aud12}Audi G \emph{et al.}, \emph{Chin. Phys.} C 36 1287 (2012); Wang \emph{et al.}, \emph{Chin. Phys.} C 36 1603 (2012).
\bibitem{Gov71}Gove N B and Martin M J \emph{At. Data Nucl. Data Tables} 10 205 (1971).
\bibitem{Led78}Lederer C M and Shireley V S  \emph{Table of Isotopes}, 7th ed. Wiley, New York, (1978).
\bibitem{Hor1}Horen D J \emph{et al.}, \emph{Phys. Lett.} B 95 27 (1980).
\bibitem{Hor2}Horen D J \emph{et al.}, \emph{Phys. Lett.} B 99 383 (1981).

\end{thebibliography}
\end{document}